\documentclass[fleqn,usenatbib]{mnras}

\usepackage[T1]{fontenc}

\DeclareRobustCommand{\VAN}[3]{#2}
\let\VANthebibliography\thebibliography
\def\thebibliography{\DeclareRobustCommand{\VAN}[3]{##3}\VANthebibliography}

\usepackage{graphicx}	% Including figure files
\usepackage{amsmath}	% Advanced maths commands
\usepackage{newtxtext,newtxmath}
\usepackage{float}
\usepackage{rotating}
\usepackage{multicol}
\usepackage{hyperref}
\usepackage{subcaption}
\usepackage{CJK}
\usepackage{color}

\title[II: Tracing composition of exoplanetary building blocks]{Seven white dwarfs with circumstellar gas discs II: Tracing the composition of exoplanetary building blocks}

\author[L. K. Rogers]{L. K. Rogers$^{1}$\thanks{E-mail: laura.rogers@ast.cam.ac.uk},
A. Bonsor$^{1}$,
S. Xu \begin{CJK*}{UTF8}{gbsn}(许\CJKfamily{bsmi}偲\CJKfamily{gbsn}艺)\end{CJK*}$^{2}$,
A. M. Buchan$^{1,3}$,
P. Dufour$^{4}$,
B. L. Klein$^{5}$,
S. Hodgkin$^{1}$, 
\newauthor M. Kissler-Patig$^{6}$,
C. Melis$^{7}$,
C. Walton$^{1,8}$,
A. Weinberger$^{9}$
\\
$^{1}$ Institute of Astronomy, University of Cambridge, Madingley Road, Cambridge CB3 0HA, UK \\
$^{2}$ Gemini Observatory, 670 N. A'ohoku Place, Hilo, HI 96720, USA \\
$^{3}$ Department of Physics, University of Warwick, Coventry CV4 7AL, UK \\
$^{4}$ D\'epartement de Physique, Universit\'e de Montr\'eal, C.P. 6128, Succ. Centre-Ville, Montr\'eal, Qu\'ebec H3C 3J7, Canada \\
$^{5}$ Department of Physics and Astronomy, University of California, Los Angeles, CA 90095-1562, USA \\
$^{6}$European Space Agency - European Space Astronomy Centre, Camino Bajo del Castillo, s/n., 28692 Villanueva de la Canada, Madrid, Spain \\
$^{7}$Center for Astrophysics and Space Sciences, University of California, San Diego, CA 92093-0424, USA \\
$^{8}$Department of Earth Sciences, ETH Zurich, Zurich, Switzerland \\
$^{9}$Earth and Planets Laboratory, Carnegie Institution for Science, 5241 Broad Branch Rd NW, Washington, DC 20015, USA
}

\date{Accepted XXX. Received YYY; in original form ZZZ}

\pubyear{2022}

\begin{document}
\label{firstpage}
\pagerange{\pageref{firstpage}--\pageref{lastpage}}
\maketitle

\begin{abstract}
%This is a simple template for authors to write new MNRAS papers.The abstract should briefly describe the aims, methods, and main results of the paper.It should be a single paragraph not more than 250 words (200 words for Letters).No references should appear in the abstract.

This second paper presents an in-depth analysis of the composition of the planetary material that has been accreted onto seven white dwarfs with circumstellar dust and gas emission discs with abundances reported in Paper I. The white dwarfs are accreting planetary bodies with a wide range of oxygen, carbon, and sulfur volatile contents, including one white dwarf that shows the most enhanced sulfur abundance seen to date. Three white dwarfs show tentative evidence (2--3\,$\sigma$) of accreting oxygen-rich material, potentially from water-rich bodies, whilst two others are accreting dry, rocky material. One white dwarf is accreting a mantle-rich fragment of a larger differentiated body, whilst two white dwarfs show an enhancement in their iron abundance and could be accreting core-rich fragments. Whilst most planetary material accreted by white dwarfs display chondritic or bulk Earth-like compositions, these observations demonstrate that core-mantle differentiation, disruptive collisions, and the accretion of core-mantle differentiated material are important. Less than one percent of polluted white dwarfs host both observable circumstellar gas and dust. It is unknown whether these systems are experiencing an early phase in the disruption and accretion of planetary bodies, or alternatively if they are accreting larger planetary bodies. From this work there is no substantial evidence for significant differences in the accreted refractory abundance ratios for those white dwarfs with or without circumstellar gas, but there is tentative evidence for those with circumstellar gas discs to be accreting more water rich material which may suggest that volatiles accrete earlier in a gas-rich phase.

\end{abstract}


%%%%%%%%%%%%%%%%%%%%%%%%%%%%%%%%%%%%%%%%%%%%%%%%%%

%%%%%%%%%%%%%%%%% BODY OF PAPER %%%%%%%%%%%%%%%%%%

\section{Introduction}

Measuring the compositions and interior structures of exoplanets is key to understanding the formation, geological history, and habitability of exoplanets across the Milky Way. Protoplanetary discs are the birth environments of planets \citep[e.g.][]{Drazkowska2023planet}. The chemical composition and structure of a protoplanetary disc are inherited from the composition of the interstellar cloud in which the star and disc collapsed, and are subsequently influenced by physical and chemical processes that mix and process the constituents of the disc \citep[e.g.][]{Visser2011chemical, Eistrup2016setting}. Planetesimals exhibit discernible variations in abundances reflective of their location within the disc in comparison to snow lines, which affects whether a species is in gaseous or condensed phase, and the mixing of reservoirs  \citep[e.g.][]{Oberg2011CO}. In the Solar System, there is a clear volatility trend where the inner system is depleted in the more volatile elements which fail to fully condense at these high inner disc temperatures  \citep[e.g.][]{Palme2003Cosmochemical, Wang2019volatility}. Therefore, the abundance patterns of planetesimals can serve as predictive indicators of their formation locations within a disc. However, measuring the bulk interior composition of planetesimals remains a challenge.

Interior modelling of exoplanets makes predictions for bulk properties of exoplanets; this is compared to the bulk density inferred from measurements of the mass and radius. However, different compositions can give rise to similar bulk densities and so degeneracies exist when inferring interior compositions \citep[e.g.][]{seager2007mass,dorn2015can}. Planetary material that pollutes the otherwise pristine atmospheres of 25--50 percent of single white dwarfs enables chemical spectroscopy of exoplanetary material \citep{zuckerman2003metal, zuckerman2010ancient, koester2014frequency, wilson2019unbiased}, and can help to break these degeneracies.

Over 1500 polluted white dwarfs are known \citep[e.g.][]{coutu2019analysis}, however, due to observational bias most of these only show evidence for pollution by calcium in optical spectra. Where multiple rock forming elements such as Ca, Mg, Si, Fe, and O are detected, most systems show planetary abundances that resemble dry, rocky, material, with refractory abundances that match chondritic meteorites \citep[e.g.][]{jura2014extrasolar,harrison2021bayesian,Trierweiler2023chondritic}. Additionally, their oxygen fugacities reveal Earth-like geochemical properties \citep{doyle2019oxygen, doyle2020extrasolar}. Therefore, dry and rocky Solar System-like planetary material appears commonplace within the local galactic population.

A handful of white dwarfs have accreted planetesimals abundant in water-ice. By comparing the abundances of rock forming elements with oxygen, those white dwarfs that have accreted planetesimals with excess oxygen are identified  \citep{farihi2011possible,farihi2013evidence, raddi2015likely, xu2016evidence, hoskin2020white,klein2021discovery}. Notably, one system was found to contain not just excess oxygen but also nitrogen, indicating the likely accretion of a Kuiper-belt-like body rich in water and nitrogen ices \citep{xu2017chemical}. Theoretical models propose that ices may sublimate and accrete before the more refractory material \citep{Malamud2021Circularization,Brouwers2023AsynchronousII}. However, an expanded sample of DAZ white dwarfs with oxygen excess calculations are required to understand whether asynchronous accretion is significant for polluted white dwarfs. 

Fragments of core-mantle differentiated bodies have been observed in the atmospheres of polluted white dwarfs, distinguished by an enrichment in siderophilic elements (iron loving elements that sink to the core) or lithophilic elements (combine readily with oxygen and tend to accumulate in the mantle of the body) \citep[e.g.][]{zuckerman2011aluminum,melis2017differentiated,hollands2018cool, harrison2018polluted}. If the white dwarfs are accreting the collisional fragments of parent bodies that underwent core-mantle differentiation, to explain these extreme observations a substantial fraction (as much as two thirds) of white dwarfs must have accreted fragments of larger core-mantle differentiated bodies \citep{bonsor2020exoplanetesimals}. This implies that the process of (iron) core formation in exo-asteroids is likely ubiquitous, and in order for differentiation to occur radiogenic heating is expected to be the dominant heat source \citep{Jura2013Al26,Curry2022prevalence,Bonsor2023rapid}.

\citet{Rogers2023sevenI} (referred to henceforth as Paper I) reported the abundances of the material that have polluted seven white dwarfs with circumstellar dust and gas discs. This paper reports an in-depth analysis of the abundances of the material that accreted onto these white dwarfs to understand the formation, collisional, and geological history of the pollutant and its parent body. The abundances used throughout this paper assume those derived using the spectroscopic white dwarf parameters, unless stated otherwise. Section \ref{Mod} presents the analytical methods utilised in this study, with details on the model, \textsc{PyllutedWD}, designed to aid the interpretation of the abundances. Section \ref{Results} presents the findings for each white dwarf, discussing the abundance ratios in comparison to Solar System objects, nearby stellar abundances, and the results from \textsc{PyllutedWD}. Section \ref{pop} presents the results on whether the abundance of the material polluting white dwarfs with detectable gaseous discs is significantly different from the material polluting white dwarfs without detectable discs in emission. Section \ref{disc} discusses the implications of the results on the chemical and geological composition of the pollutant exoplanetary material and explores the limitations of this work and the validity of the conclusions. The conclusions are presented in Section \ref{Conclusions} which summarises the findings of the paper and their broader significance for understanding the composition of exoplanetary material and the importance of key geophysical processes.

\section{Analysis Methods} \label{Mod}

\subsection{\textsc{PyllutedWD}} \label{Bayes}

In order to interpret the final abundances of the accreted planetary material reported in Table 1 in Paper I, a Bayesian framework from \citet{buchan2022planets}\footnote{\url{https://github.com/andrewmbuchan4/PyllutedWD\textunderscore Public}} is used to find the most likely explanation for the observed compositions, taking into account the differential sinking of elements in the white dwarf photosphere and the abundances of the accreted planetary material. Fundamentally, the framework focuses on the volatile content and geological history of the planetary bodies accreted by the white dwarfs \citep{harrison2018polluted, harrison2021bayesian, buchan2022planets}. \textsc{PyllutedWD} assumes that only a single planetary body is currently present in the white dwarf atmosphere. This is clearly a wide-reaching assumption as \citet{Johnson2022unusual} shows that one white dwarf may be accreting two bodies, however, as discussed in \citet{harrison2018polluted} and \citet{turner2020modelling}, it is likely that the mass currently in the atmosphere is dominated by the most massive, single body accreted and so the assumption of a single body dominating the abundances is reasonable. Three main processes are considered that dominate changes to the planetary composition. First, the composition of the initial material available for planet formation may differ from the Solar System. Second, the planetary bodies may be depleted in volatiles due to the high temperatures experienced during formation or subsequent evolution. Third, if large-scale melting occurs, the segregation of the iron melt leads to the formation of a core and a mantle under the influence of the internal pressure and oxygen fugacity. Subsequent collisions or other processing that leads to fragmentation can change the relative amount of core or mantle material that is accreted by the white dwarf. 

\textsc{PyllutedWD} uses a Bayesian framework to compare the evidence for a basic primitive model, in terms of its ability to explain the abundance and abundance upper limits of a white dwarf pollutant, in comparison to a range of more complex models. The basic primitive model is only dependent on the initial composition in which the planetary material formed and sinking effects in the atmosphere of the white dwarf. The basic free parameters are listed below, with the additional more complex parameters listed as optional:
\begin{itemize}
    \item `{Initial Composition':} the body accreted by the white dwarf could have formed in planetary systems with a range of initial compositions. The compositions of nearby stars is a good proxy for this range of initial compositions \citep{brewer2016c}. Most cases find that the white dwarf is equally likely to have accreted material that started with a wide range of initial compositions. 
    \item `{Pollution level}': the mole fraction of atoms in the white dwarf's convection zone which are metals (and therefore assumed to be pollutants), M$_{Z}$/M$_{CVZ}$.
    \item `{Accretion event time-scale':} the length of time over which material is accreted on to the white dwarf. 
    \item `{Time since accretion on to the white dwarf'}: the time elapsed since the start of the accretion event such that if this is greater than the accretion event timescale, accretion has ceased prior to observation.   
    \item `\textit{Temperature}' (optional complex parameter): the depletion of volatiles is linked to the temperature (pressure) conditions at a certain location in the protoplanetary disc. This assumes that patterns of volatile depletion occur due to the incomplete condensation of hot gas in the inner regions of a protoplanetary disc. It is also possible that heating occurred after planetesimal formation via collisions, impacts, or for small planetesimals during the late stages of stellar evolution \citep{Li2024PostMS}.
    \item `\textit{Feeding Zone Size}' (optional complex parameter): the abundances of some planetary material are best explained by a mixture of material that experienced different temperatures during formation. This would occur in reality when planetary bodies accrete material from a range of radial locations, or their feeding zone.
    \item `\textit{Fragment Core Fraction}' (optional complex parameter): the white dwarf may have accreted material that is dominated by the core or the mantle of a larger planetary body. The core number fraction and mass fraction are both outputted.
    \item `\textit{Core-mantle equilibration pressure}' and `\textit{Core-mantle oxygen fugacity}' (optional complex parameter): these parameters determine the composition of the core and mantle based on the conditions that were present during core-mantle differentiation, notably the pressure and the oxygen fugacity (i.e., its availability to oxidise other metals). The composition of the core or mantle, most notably the Cr, Si, and Ni content, is constrained based on a model for elemental partitioning during core formation, this model allows the core composition to differ from Earth's core and has two free parameters: `\textit{core-mantle equilibration pressure}' and `\textit{oxygen fugacity}' at the time of core formation (note this is different to oxygen fugacity of the final planetary body, as considered in \citealt{doyle2019oxygen}). The mid-mantle pressure and oxygen fugacity of the Earth as found from \citet{buchan2022planets} are 45 GPa and $-$1.3, where the oxygen fugacity is reported in log units relative to the Iron-W{\"u}stite reaction ($\Delta$IW). An asteroid sized body, with significantly lower pressure than Earth, would have a core with higher Ni/Fe and a lower Cr/Fe and Si/Fe than the Earth's core, 

\end{itemize}

The Bayesian framework compares the basic primitive model with no additional free parameters included (null hypothesis, M$_0$), with more complex models which incorporate additional free parameters, (alternative hypothesis, M$_1$). It is assessed whether the introduction of additional free parameters results in an improved Bayesian evidence using the Bayes factor. This compares the likelihood of the data given the alternative hypothesis, M$_1$, to the likelihood of the data given the null hypothesis, M$_0$. A Bayes factor of $>$ 1 implies evidence for the alternate hypothesis, and a Bayes factor of $>$ 100 implies strong evidence for an alternate hypothesis. Additionally, this is converted to a frequentist statistic (N\,$\sigma$) to also show the significance of the model. 

There are three stages of accretion and the Bayesian framework finds the most likely stage. Build up stage - when accretion has just started so the observed abundances correspond closely to the actual abundances of the pollutant,
\begin{equation}
    \frac{nX(A)_{\textrm{\,par}}}{nX(B)_{\textrm{\,par}}} = \frac{nX(A)_{\textrm{\,WD}}}{nX(B)_{\textrm{\,WD}}},
    \label{eq:1}
\end{equation}
where $nX(A)_{\textrm{\,par}}$ and $nX(B)_{\textrm{\,par}}$ are the abundances for element A and B respectively of the parent body before being accreted by the white dwarf, and $nX(A)_{\textrm{\,WD}}$ and $nX(B)_{\textrm{\,WD}}$ are the derived number abundances for element A and B from the observations of the photosphere of the white dwarf \citep{koester2009accretion,harrison2018polluted}. Steady state stage - when accretion is ongoing, and has been going on for long enough that the abundances in the white dwarf atmosphere have reached an equilibrium. For this phase the abundances derived for the accreted material need to be modified to consider gravitational settling for the elements in the white dwarf photosphere,
\begin{equation}
    \frac{nX(A)_{\textrm{\,par}}}{nX(B)_{\textrm{\,par}}} = \frac{\tau _{B}}{\tau _{A}}\frac{X(A)_{\textrm{\,WD}}}{X(B)_{\textrm{\,WD}}},
    \label{eq:2}
\end{equation}
where $\tau _{A}$ and $\tau _{B}$ are the diffusion time-scales of element A and B respectively through the white dwarf's photosphere. Declining phase - after the parent body has been fully accreted by the white dwarf, the abundances decrease exponentially with the decay factors depending on the time since accretion ceased, \textit{t},
\begin{equation}
    \frac{nX(A)_{\textrm{\,par}}}{nX(B)_{\textrm{\,par}}} \propto \frac{nX(A)_{\textrm{\,WD}}}{nX(B)_{\textrm{\,WD}}} \frac{e^{-t/\tau _{B}}}{e^{-t/\tau _{A}}}.
    \label{eq:3}
\end{equation}

As all the objects have circumstellar gas and dust reservoirs, accretion is likely ongoing and dominated by one singular body. This is almost certainly occurring for the hot DA white dwarfs, with settling timescales on the order of days \citep{koester2009accretion}\footnote{ Using the 2020 overshoot values from:  \url{http://www1.astrophysik.uni-kiel.de/~koester/astrophysics/astrophysics.html}}. The most likely phase of accretion is used as a consistency check to ensure that the model doesn't find the most likely phase of accretion being in the declining phase, which is deemed unlikely due to the presence of the circumstellar reservoirs. The abundances and output of the Bayesian framework are discussed in the results section below.

\subsection{Oxygen Budget} \label{Oxygen}

To determine whether the pollutant is water rich, has any iron in metallic form, or is dry, rocky material, oxygen budgeting is done following the method used by \citet{klein2010chemical}. O is assigned to the oxides: MgO, SiO$_2$, Al$_2$O$_3$, CaO, and FeO. If there is excess oxygen, this was likely in the form of water in the parent body. The abundances are corrected for sinking assuming the accretion is occurring in the steady state for DAs or build up and steady state for DBs. If there is not a measured abundance for one or more of the major rock forming elements (Mg, Si, Al, Ca and Fe) then the upper limit abundance of that element is used in the calculation. The oxidation state of Fe is unknown and Fe could be in a number of forms (or a combination) including: FeO, Fe$_2$O$_3$, or metallic Fe. Therefore, oxygen budgeting is used to determine if there is excess oxygen rather than to explore the exact water content of a body. The errors on the oxygen excess are calculated using Monte Carlo to sample the errors on the abundance. This calculation does not consider carbon ices, however, as a test, including CO in the calculations changes the oxygen excess fraction by less than 0.002 (0.2 per cent). 

The Bayesian framework, \textsc{PyllutedWD}, also calculates an oxygen excess \citep{Brouwers2023AsynchronousII}. This samples from the errors on the abundances (for those elements without detections it used the median of the posterior distribution for that element) and from the posterior distribution on the phase of accretion. The significance of the oxygen excess is outputted and used as a consistency check with the calculations reported above. 

\begin{figure*}
%\captionsetup[subfigure]{font=small} if you like to change caption style
     \begin{subfigure}[b]{0.49\textwidth}
         \includegraphics[width=\columnwidth]{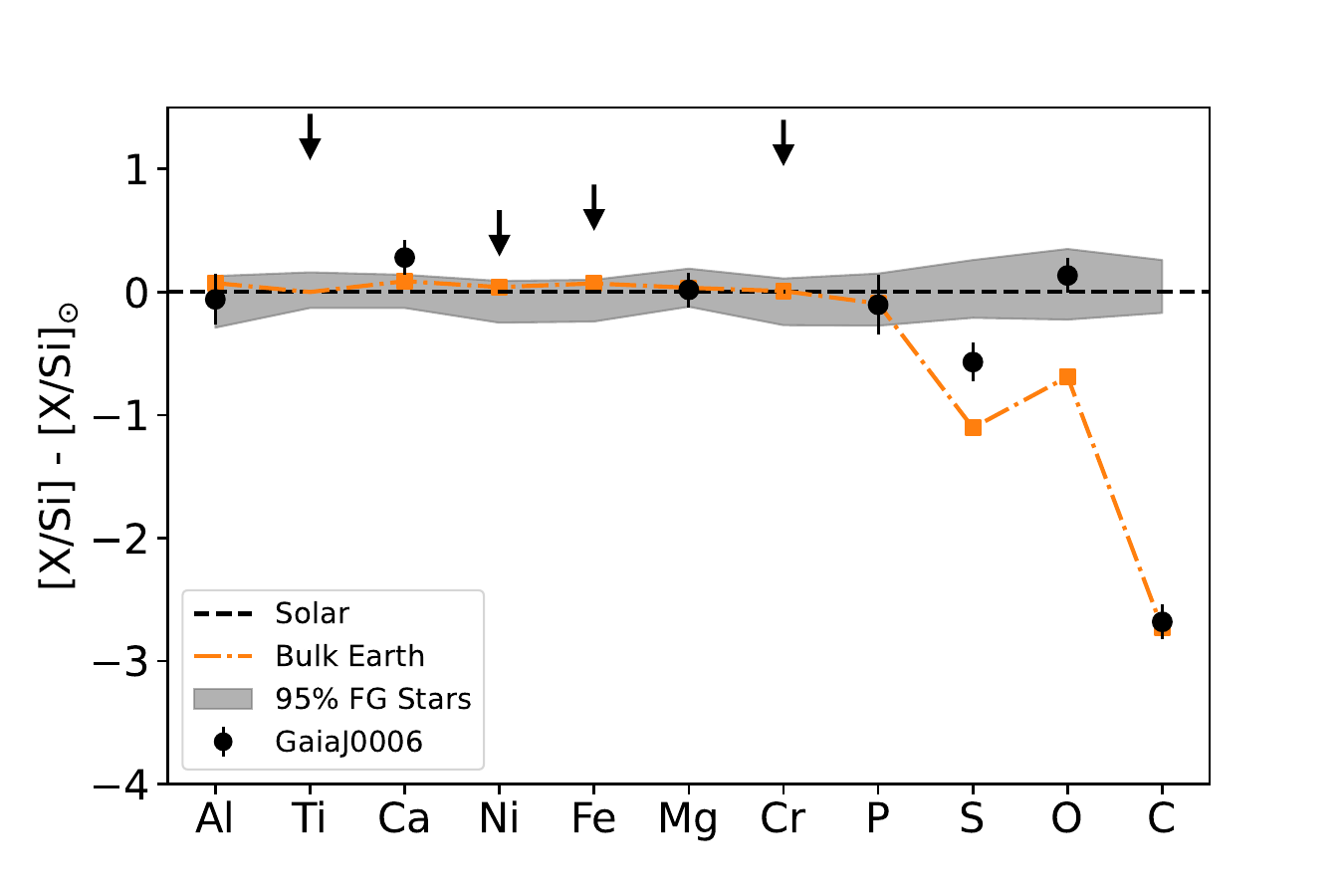}
         \caption[]{Gaia\,J0006+2858} % <---
         \label{fig:GaiaJ0006-Ab}
     \end{subfigure}
     \hfill
     \begin{subfigure}[b]{0.49\textwidth}
         \includegraphics[width=\columnwidth]{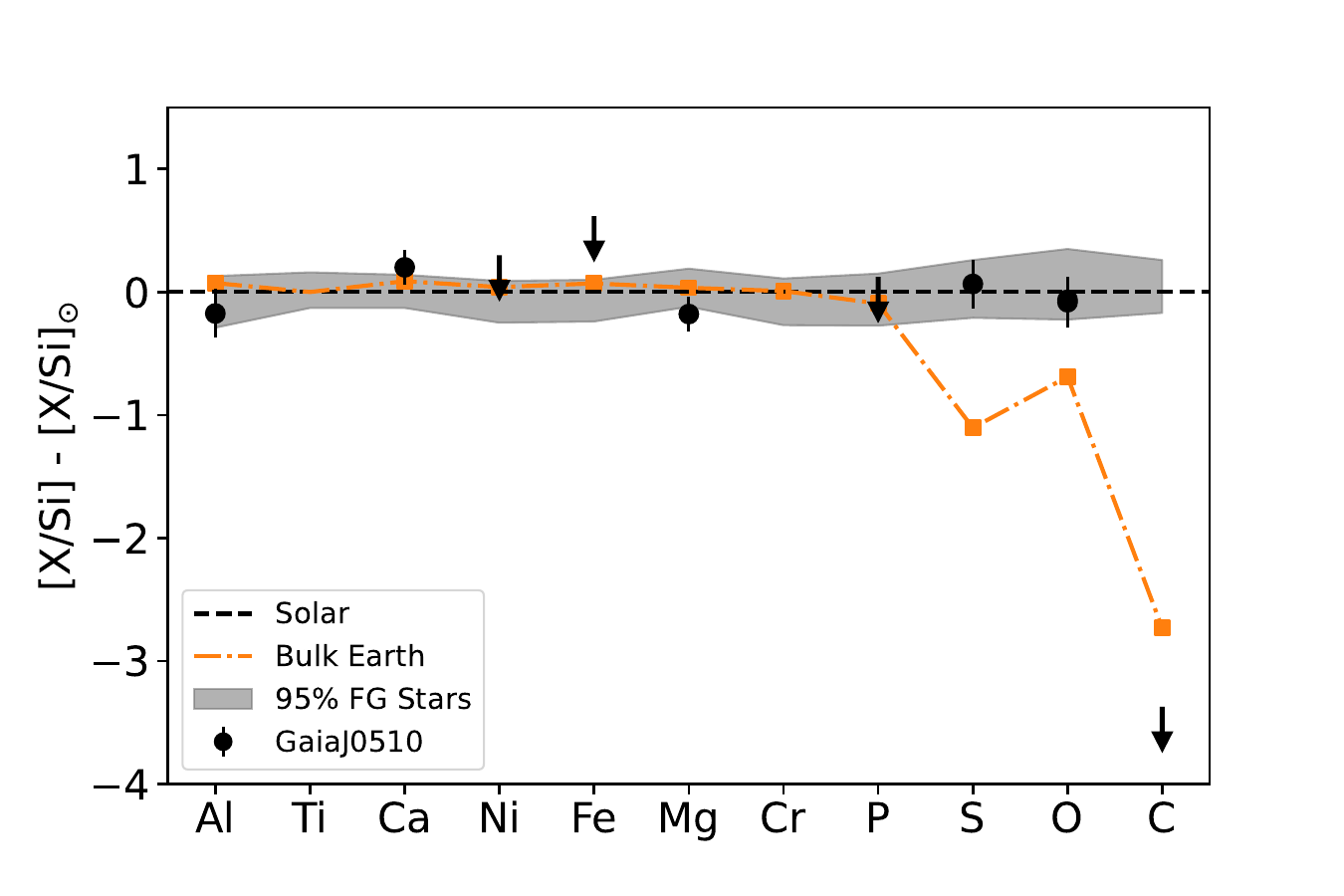}
         \caption[]{Gaia\,J0510+2315} % <---
         \label{fig:GaiaJ0510-Ab}
     \end{subfigure}

     \vskip\baselineskip
     \begin{subfigure}[b]{0.49\textwidth}
         \includegraphics[width=\columnwidth]{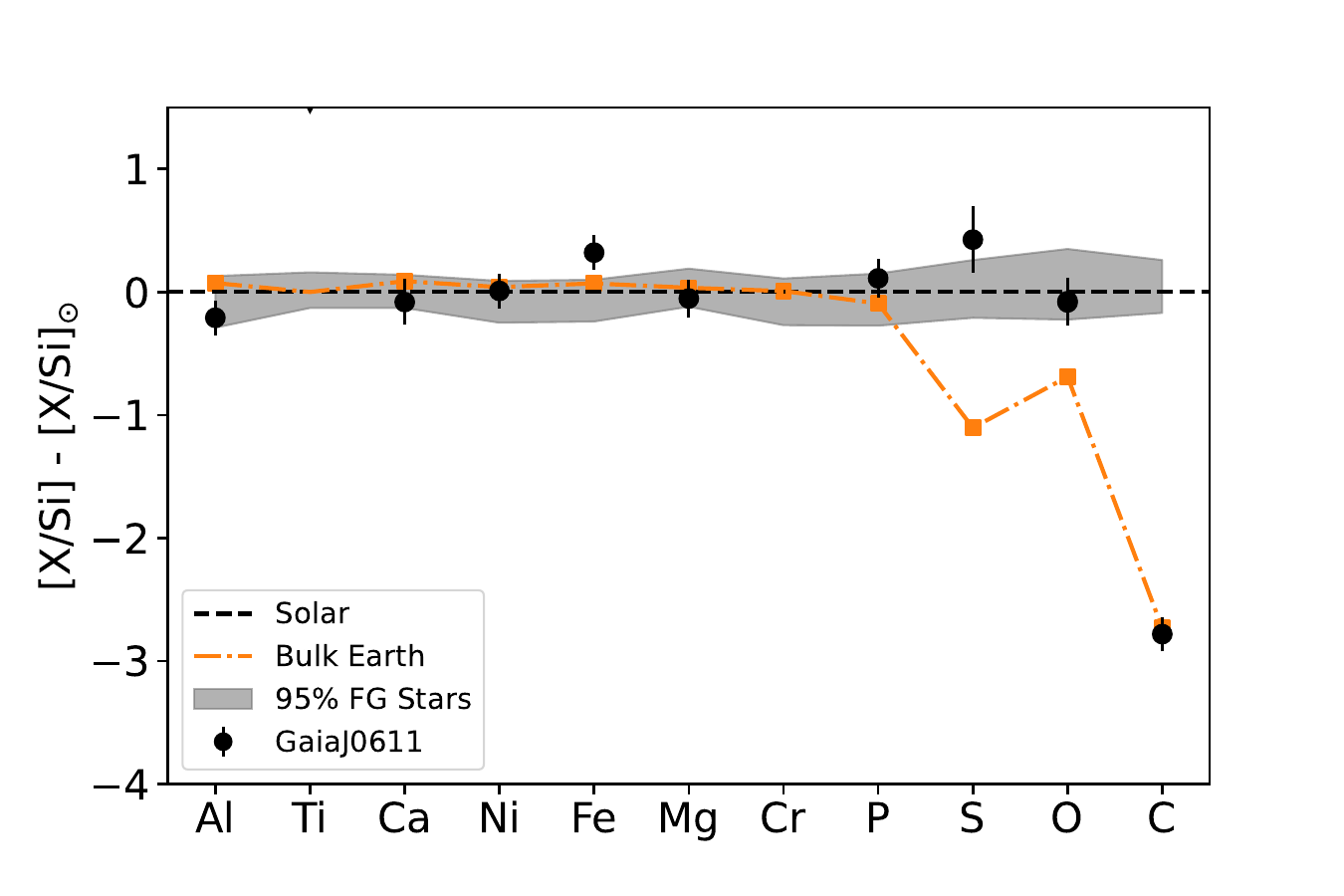}
         \caption[]{Gaia\,J0611$-$6931} % <---
         \label{fig:GaiaJ0611-Ab}
     \end{subfigure}
     \hfill
     \begin{subfigure}[b]{0.49\textwidth}
         \centering
         \includegraphics[width=\columnwidth]{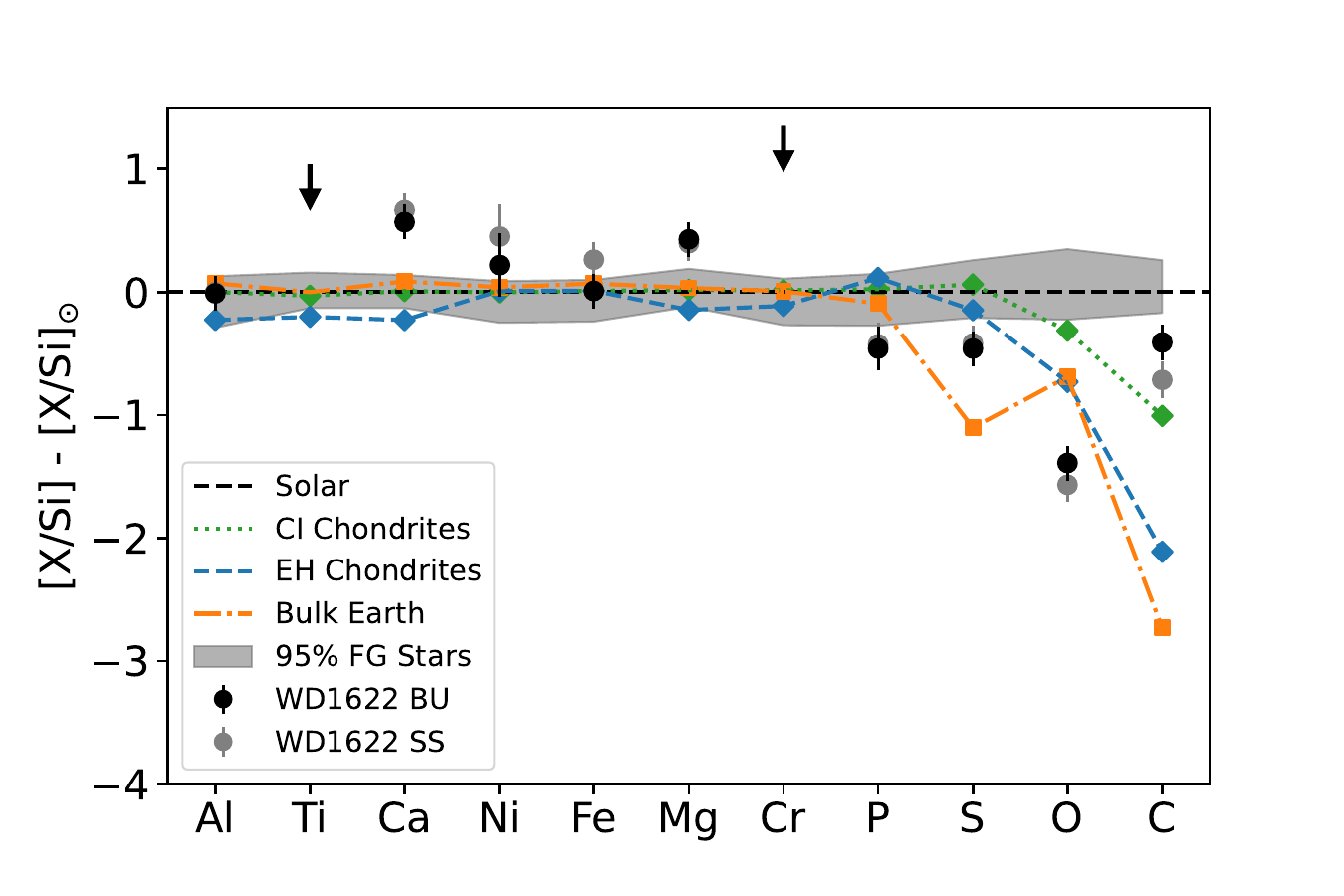}
         \caption[]{WD\,1622+587} % <---
         \label{fig:WD1622-Ab}
     \end{subfigure}
        \caption{The logarithmic number ratio of elemental abundances in the parent body relative to Si, normalised to solar (black dashed line) for the four white dwarfs with optical and ultraviolet data. The grey shaded region shows the 95 percent range of the abundance ratios of main sequence FG stars from the Hypatia catalog \citep{hinkel2014stellar}. For the DA white dwarfs (Gaia\,J0006+2858, Gaia\,J0510+2315, and Gaia\,J0611$-$6931) the abundances are adjusted for sinking assuming steady state given the short settling timescales (Table \ref{tab:sinking-timescales}). For WD\,1622+587, a DB white dwarf, the abundances are plotted assuming both build-up (WD1622 BU) and steady state phase (WD1622 SS). Upper limits are plotted as arrows and are measured from the base of the arrow, and for some elements lie outside of the range of the plot.} % <---
        \label{fig:aenb}
\end{figure*}

\section{Results from individual systems}  \label{Results}

This section is dedicated to the interpretation of the observed abundances for the planetary material that were accreted by each white dwarf, whilst Section \ref{pop} provides an overarching interpretation of the entire population. Throughout this analysis, the abundances obtained based on the spectroscopically derived stellar parameters are used. It is worth noting, as highlighted in Paper I, when considering abundance ratios rather than absolute abundances, the two sets of measurements agree within a 1\,$\sigma$ error. Due to the optical-ultraviolet discrepancy, silicon serves as the reference element for abundance ratios as it is measured in both datasets; whether the optical or ultraviolet silicon abundance is used depends on whether the element being considered was measured using the optical or ultraviolet. Abundances are by number unless stated otherwise. The abundance ratios for each target inform our understanding of the history and state of the planetary material accreted, and these interpretations are presented first, followed by more detailed insights derived from the Bayesian framework discussed in Section \ref{Bayes}. When comparing to stellar compositions the Hypatia catalogue is used \citep{hinkel2014stellar}, except for when using results from \textsc{PyllutedWD} which uses the catalogue from \citet{brewer2016c}. The choice of stellar catalogue does not significantly alter the results.

\subsection{Gaia\,J0006+2858} \label{GaiaJ0006}

% primitive and oxygen excess

\begin{figure*}
	\includegraphics[width=\columnwidth]{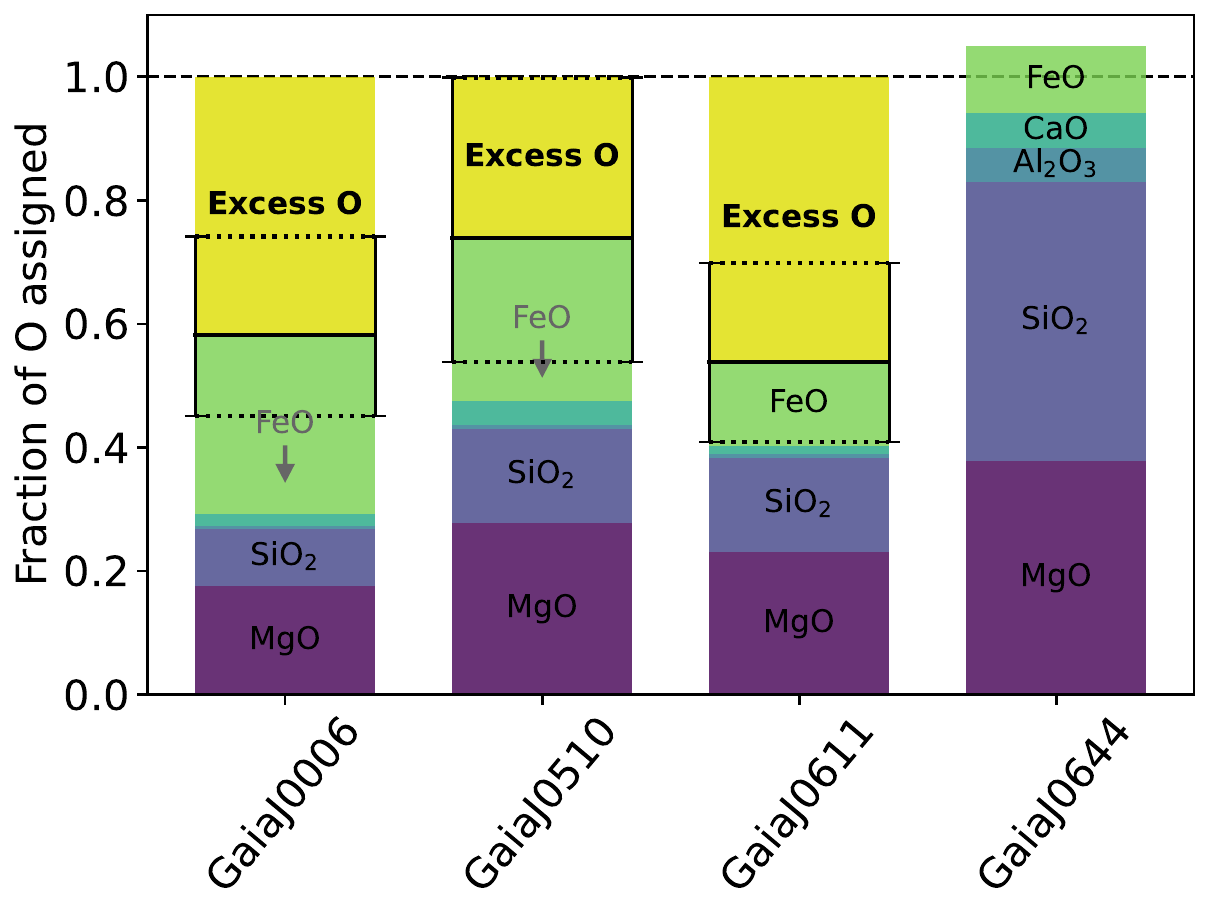}
    \caption{The likely division of the observed oxygen abundance between metal oxides (MgO, SiO$_2$, Al$_2$O$_3$, CaO, and FeO) and water-ice (yellow), for Gaia\,J0006+2858, Gaia\,J0510+2315, Gaia\,J0611$-$6931, and Gaia\,0644$-$0352. The approach is described in Section \ref{Oxygen}, using abundances adjusted for sinking in steady state. The error bars on the oxygen excess are 1\,$\sigma$ errors calculated using Monte Carlo to sample the errors on the measured elemental abundances. The Fe for Gaia\,J0006+2858 and  Gaia\,J0510+2315 are upper limits and so the FeO is labelled as an upper limit with an arrow.}
    \label{fig:O-ex}
\end{figure*}

The planetary material accreted by Gaia\,J0006+2858 has close to solar Al/Si, Ca/Si, Mg/Si, P/Si and O/Si, but depleted S/Si and C/Si (see Fig.\,\ref{fig:GaiaJ0006-Ab}). Solar (or close to) abundance ratios of O/Si are seen in Kuiper belt objects from the cold outer regions of the Solar System. The C/Si abundance ratio is depleted in comparison to solar and in line with the abundance ratio of rocky material. The material is likely rocky material with water. Assuming steady state accretion, the derived oxygen budget described in Sec \ref{Oxygen} (Table\,\ref{tab:WDs-ab-O}) is sufficiently large that extra oxygen is indicated, over and above that which can be accounted for in metal oxides, see Fig.\,\ref{fig:O-ex}. The upper limit on the Fe abundance was used in this calculation, therefore the oxygen excess is a lower limit. The derived fractional oxygen excess is $>$\,0.41 (41 per cent), considering only FeO, or more conservatively considering Fe$_2$O$_3$, $>$\,0.27 (27 per cent), therefore, even if all the Fe that could have been hidden below the detection limit was put into FeO or Fe$_2$O$_3$ there would still be an oxygen excess.

\textsc{PyllutedWD}, described in Sec \ref{Bayes} also finds that Gaia\,J0006+2858 is most likely to be accreting `primitive' material in steady state, where the abundances depend only on the initial composition of the stellar nebula and the white dwarf accretion and sinking parameters, with a $2.5\sigma$ oxygen excess.

\begin{table*}
	\centering
	\footnotesize
	\caption{Number abundances (log n(Z)/n(H(e)) of the material polluting the white dwarfs to be used for the oxygen excess calculation. Abundances are taken from Table 1 in Paper I using the spectroscopically derived values and ultraviolet data, where available. If the O excess is equal to 0 then the oxygen budget is balanced with respect to the oxides, if it is negative then there is not enough oxygen present to oxidise the major rock forming elements, and metallic forms likely exist. The upper limits of Fe are used in the calculation of O excess for Gaia\,J0006 and Gaia\,J0510 and so the O excess is a lower limit. 
	% Al, Ti, Ca, Ni, Fe, Cr, Si, Na, O
	}
	\label{tab:WDs-ab-O}
	\begin{tabular}{lccccccccc} % four columns, alignment for each
		\hline
		[X/H(e)] & {Gaia\,J0006} & {Gaia\,J0510} & {Gaia\,J0611} & {Gaia\,J0644} & {WD\,1622} \\
		\hline	
		
		O & $-$4.48\,$\pm$\,0.10 & $-$4.98\,$\pm$\,0.12 & $-$4.28\,$\pm$\,0.10  &  $-$5.17\,$\pm$\,0.13 & $-$5.39\,$\pm$\,0.10 \\ 

		Mg & $-$4.95\,$\pm$\,0.10 &  $-$5.23\,$\pm$\,0.10 & $-$4.61\,$\pm$\,0.11 & $-$5.73\,$\pm$\,0.10 & $-$4.76\,$\pm$\,0.10  \\
  
	    Al & $-$6.50\,$\pm$\,0.18 & $-$7.04\,$\pm$\,0.10 & $-$6.46\,$\pm$\,0.10 & $-$6.76\,$\pm$\,0.11 & $-$6.28\,$\pm$\,0.10 \\

        Si & $-$5.48\,$\pm$\,0.10 & $-$5.85\,$\pm$\,0.17 & $-$5.22\,$\pm$\,0.10 & $-$5.97\,$\pm$\,0.10 & $-$5.20\,$\pm$\,0.10 \\
		
        Ca & $-$6.17\,$\pm$\,0.10 & $-$6.31\,$\pm$\,0.10 & $-$6.08\,$\pm$\,0.15 & $-$6.70\,$\pm$\,0.10 & $-$5.85\,$\pm$\,0.10 \\
		
		Fe & $<-$5.00 & $<-$5.60 & $-$5.23\,$\pm$\,0.10 & $-$6.51\,$\pm$\,0.10 & $-$5.26\,$\pm$\,0.10 \\

        \hline
        O Excess SS (FeO) & $>0.41\substack{+0.13 \\ -0.16}$ & $>0.25\substack{+0.20 \\ -0.26}$ & $0.45\substack{+0.13 \\ -0.16}$ & $-0.07\substack{+0.30 \\ -0.41}$ & $-13.8\substack{+3.4 \\ -4.4}$  \vspace{1mm} \\ 
        
        O Excess SS (Fe$_2$O$_3$) & $>0.27\substack{+0.15 \\ -0.20}$ & $>0.11\substack{+0.23 \\ -0.30}$ & $0.38\substack{+0.14 \\ -0.19}$ & $-0.13\substack{+0.31 \\ -0.43}$ & $-15.5\substack{+3.8 \\ -5.0}$ \vspace{1mm} \\
        
        O Excess BU (FeO) & - & - & - & $0.28\substack{+0.20 \\ -0.28}$ & $-8.4\substack{+2.2 \\ -2.9}$ \vspace{1mm}  \\
        
        O Excess BU (Fe$_2$O$_3$) & - & - & - &  $0.26\substack{+0.20 \\ -0.29}$ & $-9.1\substack{+2.3 \\ -3.0}$ \vspace{1mm}  \\
		\hline
	\end{tabular}
\end{table*}

\subsection{Gaia\,J0347+1624} \label{GaiaJ0347}

The story for Gaia\,J0347+1624 is difficult to disentangle, as only Al and Mg have reliable detections, given the contribution of the interstellar medium to the Ca line. As Al was detected in the near-ultraviolet and Mg in the optical, care must be taken in interpreting the low [Al/Mg] ratio of $-$0.43 compared to solar (assuming steady state), as the optical-ultraviolet discrepancy could be affecting the abundance ratio. Less than 1 percent of the Hypatia catalogue FG main sequence stars have an Al/Mg ratio this low \citep{hinkel2014stellar}. If the Al/Mg ratio of the planetary material is truly sub-solar then Gaia\,J0347+1624 adds to a number of polluted white dwarfs with low Al/Mg ratios, for example, WD\,1929+012 \citep{gansicke2012chemical}. 

\subsection{Gaia\,J0510+2315} \label{GaiaJ0510}

% Primitive with O excess than is 2 sigma ish. 

The abundance ratios of the exoplanetary body that polluted Gaia\,J0510+2315 has Al/Si, Ca/Si, Mg/Si, S/Si and O/Si that are close to solar and within the abundance range of nearby stars (see Fig.\,\ref{fig:GaiaJ0510-Ab}). The oxygen abundance is consistent with the accretion of a volatile rich body and has a similar abundance ratio to solar, rather than depleted in volatiles, which is the case for the rocky material typically accreted by white dwarfs. The C/Si upper limit lies below that expected from bulk Earth and is therefore constraining. As reported for Gaia\,J0006+2858 in Section \ref{GaiaJ0006}, the oxygen budget was calculated using the abundances reported in Table \ref{tab:WDs-ab-O}. Based on the available data, Fe is not detected in the atmosphere of this white dwarf and so the upper limit is used for the oxygen budget, and therefore, as with Gaia\,J0006+2858 the fractional oxygen excess is a lower limit. Both oxidisation strategies of Fe show an oxygen excess: $>$\,0.25 for FeO and $>$\,0.11 for Fe$_2$O$_3$, suggesting the body may have had a contribution from water. Gaia\,J0510+2315 is a DA, so hydrogen from an accreted body cannot be spectroscopically distinguished from the hydrogen envelope of the white dwarf. The oxygen circumstellar gaseous emission features are particularly strong implying there could also be volatile-rich circumstellar material. Full radiative transfer calculations such as those in \citet{gansicke2019accretion} and \citet{steele2021characterization} are required to investigate whether the circumstellar gas is rich in oxygen as the photosphere appears to be; this is beyond the scope of this paper.

\textsc{PyllutedWD} finds that the most likely explanation for the observed abundances is the basic model where, like Gaia\,J0006+2858, Gaia\,J0510+2315 is accreting `primitive' material in steady state. Using the \citet{Brouwers2023AsynchronousII} oxygen excess calculations included in \textsc{PyllutedWD}, there is evidence of an oxygen excess at 2.6\,$\sigma$.

\subsection{Gaia\,J0611$-$6931} \label{GaiaJ0611}

% Primitive with O excess than is 2 sigma ish. 

Gaia\,J0611$-$6931 has accreted a planetary body with approximately solar abundance ratios of Al/Si, Ca/Si, Ni/Si, Mg/Si, P/Si, S/Si, and O/Si (as seen in Fig.\,\ref{fig:GaiaJ0611-Ab}). The Fe/Si ratio is enhanced by 0.31\,dex compared with a solar Fe/Si ratio assuming steady state (inconsistent with a solar ratio at the 2.3\,$\sigma$ level). Less than 0.1 percent of Hypatia catalogue FG stars have a Fe/Si ratio greater than 0.31\,dex \citep{hinkel2014stellar}, so it is unlikely that the progenitor star and planetesimals formed in an iron rich environment, and this may point towards the accretion of a fragment of material that is enriched in iron. The oxygen abundance is approximately solar and oxygen budgeting revealed that this body was likely oxygen and therefore water rich. Both of the oxidisation strategies for Fe show a significant fractional oxygen excess: 0.45 for FeO and 0.38 for Fe$_2$O$_3$. Additionally, some Fe could be metallic and so would not contribute to the budget. 

The most likely \textsc{PyllutedWD} model invoked to explain the observed abundances is the basic model where, like Gaia\,J0006+2858 and Gaia\,J0510+2315, it is accreting `primitive' material in steady state. The Fe enhancement is not high enough to reject the primitive model, given the abundance errors and the penalty of including additional parameters in more complex models in the context of the Bayesian framework. Using the \citet{Brouwers2023AsynchronousII} oxygen excess calculations included in \textsc{PyllutedWD}, there is evidence that the material accreted onto Gaia\,J0611$-$6931 was oxygen-rich, and had a 2.1\,$\sigma$ oxygen excess.

\subsection{Gaia\,J0644$-$0352} \label{WD0644}

% Differentiation and mantle rich fragment, and heating to explain the O depletion. 
% But photometric solution puts it into declining phase. 
% The model exhibits a slight preference for Earth-like differentiation, with the mode values of pressure and oxygen fugacity being 42 GPa and IW - 2.1 respectively. The errors are large, however.

\begin{figure*}
\centering
\subfloat[]{
  \includegraphics[width=0.58\textwidth]{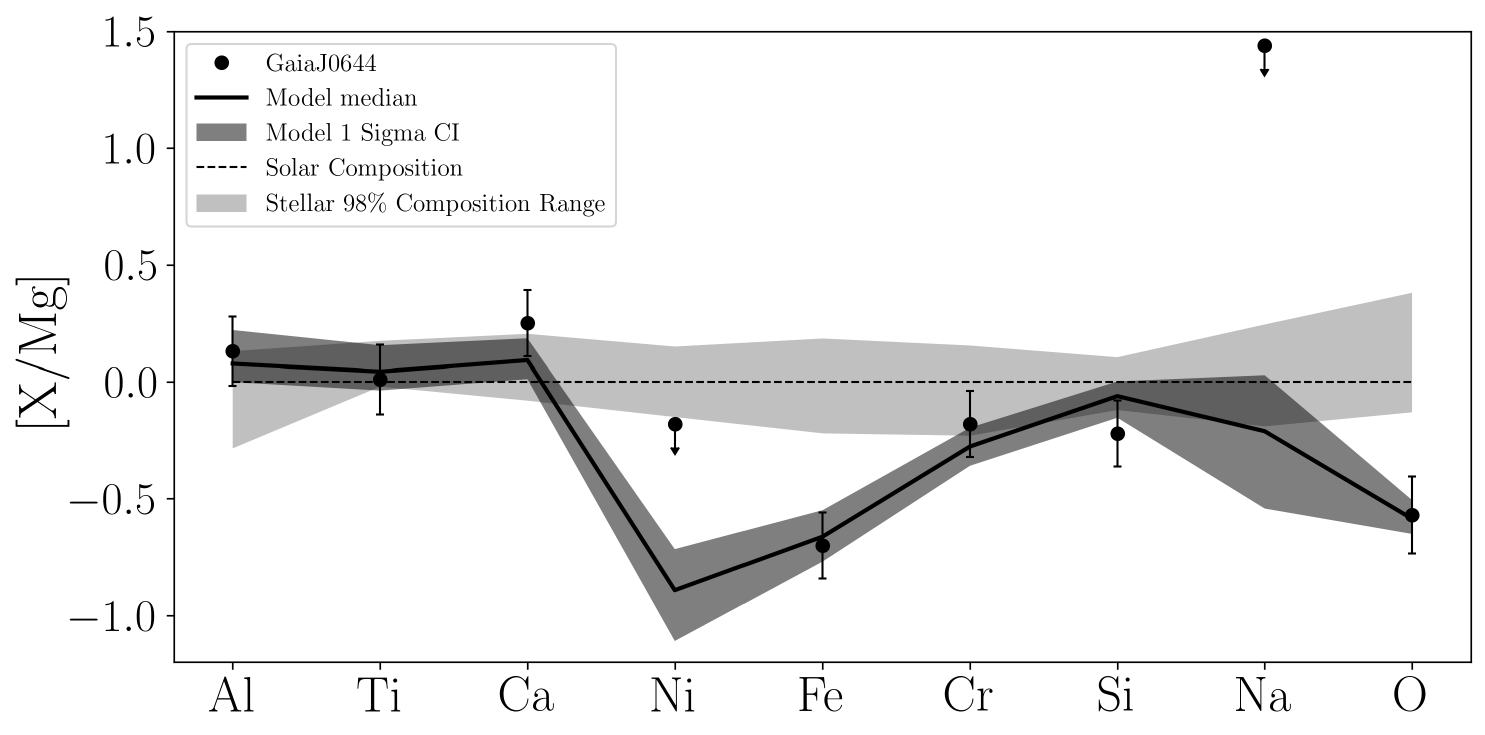}
  \label{fig:GaiaJ0644-Model-a}
}
\hspace{2mm}
\subfloat[]{
  \includegraphics[width=0.38\textwidth]{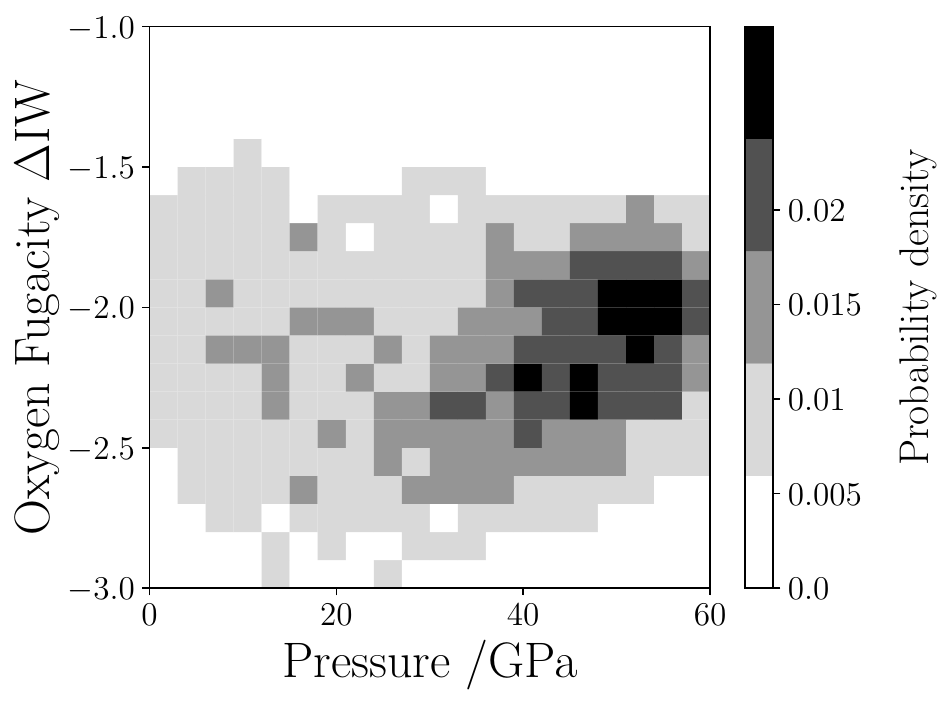}
  \label{fig:GaiaJ0644-Model-b}
}

\caption{(a) The logarithmic number ratio of elemental abundances in the pollutant material relative to Mg, normalised to solar (dashed line). Observed abundances for the white dwarf, Gaia\,J0644$-$0352, are shown as black points with 1$\sigma$ errors, assuming build-up phase as sinking effects are included in the model. The black solid line represents the maximum likelihood median model with heating and differentiation included (a `mantle-rich' model) and a 1$\sigma$ error on this model as the shaded dark grey region. The shaded light grey region indicates the range of abundances seen in nearby stars using the catalogue from \citet{brewer2016c}. (b) Posterior distributions of pressure and oxygen fugacity for the parent body to the fragment that polluted Gaia\,J0644$-$0352. The modal values of pressure and oxygen fugacity are 42\,GPa and $\Delta$IW $-$2.1 respectively, which compare to that of the mid-mantle of the Earth, 45\,GPa and $-$1.3.}
\end{figure*}

Gaia\,J0644$-$0352 has solar abundances of Al, Ti, Ca, Cr, and Si, when compared to Mg, but has sub-solar O and Fe, as shown in Fig.\,\ref{fig:GaiaJ0644-Model-a}. The oxygen abundance is depleted by $>$\,3\,$\sigma$ in comparison to a solar abundance indicating the accretion of rocky material. Indeed, oxygen budgeting calculations using the abundances reported in Table \ref{tab:WDs-ab-O} found that there are sufficient abundances of the main rock forming elements for oxygen to be carried entirely in metal oxides, as shown by Fig.\,\ref{fig:O-ex}. Therefore, this white dwarf is accreting dry, and rocky material. The low Fe abundance could be due to the accretion of a mantle-rich fragment which would naturally explain a depletion in iron, and the sub-solar upper limit of Ni. Indeed, comparison of the abundance ratios of the material accreting on to Gaia\,J0644$-$0352 with abundance ratios in the Earth's mantle shows a good agreement \citep{mcdonough2003compositional}. Therefore, Gaia\,J0644$-$0352 is likely accreting a mantle fragment of a larger differentiated parent body.

From \textsc{PyllutedWD} the most likely explanation for the observed abundances (highest Bayesian evidence) is the accretion of volatile depleted mantle-rich material. A high temperature that depletes the material in volatiles, in this case water (or O), is preferred over a primitive model with a high Bayes factor (8.9$\times 10^{8}$ or 6.7\,$\sigma$) giving strong evidence for heating. Additionally, the Bayesian model that invokes core-mantle differentiation is favoured over a model without differentiation with a high Bayes factor (110 or 3.5\,$\sigma$) giving strong evidence for the accretion of a core-mantle differentiated fragment. Therefore, a model in which Gaia\,J0644$-$0352 accreted volatile-poor and mantle-rich material is a good fit to the data, as shown by the grey median model in Fig.\,\ref{fig:GaiaJ0644-Model-a}. Sampling the posterior distribution of the fragment core mass fraction gives a median value of 0.05\,$^{+0.04}_{-0.03}$ (the core mass fraction of the Earth is 0.325), meaning this white dwarf is accreting a mantle rich fragment. The modal values of pressure and oxygen fugacity are 42\,GPa and $\Delta$IW $-$2.1 respectively, which compare to that of the Earth's mid-mantle, 45\,GPa and $-$1.3 respectively, as shown in Fig.\,\ref{fig:GaiaJ0644-Model-b}. This may imply that Gaia\,J0644$-$0352 is accreting a fragment of a core-mantle differentiated body that was planetary sized, rather than asteroid sized. It should be noted however that the posterior distribution on the pressure overlapped into the low pressure region and consequently there is a non-negligible probability that this was a smaller parent body.

\subsection{WD\,1622+587} \label{WD1622}

% Optical - invokes heating to 10.4 sigma. 
% UV - extreme heating and oxygen depletion.
% Not enough oxygen to fully oxidise everything. Oxygen depletion. 

The Al/Si, Ni/Si, and Fe/Si abundance ratios assuming build-up phase of the exoplanetary body that polluted WD\,1622+587 are consistent with solar and compositions seen in nearby main sequence FG stars \citep{hinkel2014stellar}, as seen in Fig.\,\ref{fig:WD1622-Ab}. The Ca/Si, Mg/Si are super-solar, and the P/Si, S/Si, O/Si and C/Si are sub-solar. The oxygen abundance suggests that the accreted body is depleted in oxygen, and the O/Si ratio is an order of magnitude below that seen in Bulk Earth like material. The oxygen budgeting reveals that there is insufficient oxygen to oxidise the rock forming elements, as shown in Table\,\ref{tab:WDs-ab-O}, in fact, the oxygen is so low that between 8-15 times more oxygen would be required to form the metal oxides and so it is likely that the body that accreted onto this white dwarf was very reduced. 

Fig.\,\ref{fig:WD1622-Ab} shows the abundance patterns of EH chondrites, one of the main enstatite groups, and CI chondrites for comparison. Enstatite chondrites are some of the most reduced bodies in the meteorite record, and even this composition does not match the abundances of WD\,1622+587. In order to explain the depleted oxygen, the body may have contained: metallic Fe and Ni; compounds such as iron sulphide (FeS); sulphides of typically lithophilic elements (e.g. oldhamite CaS, or niningerite MgS); or carbides (e.g. moissanite SiC or magnesium carbide MgC). It is also possible that carbides of Ca, Mg, Ti, and Al could form if the conditions were sufficiently reduced given the large amounts of these elements and the carbon abundance, but this is rarely seen in nature.  The carbon to silicon abundance ratio is close to being solar and significantly in excess compared to material similar to Bulk Earth, so there may be a significant amount of carbon to form these carbides. However, it is known that hot DB white dwarfs ($>$\,20\,000\,K) often show excess carbon that may be attributed to carbon winds which can maintain carbon in the atmosphere of the white dwarf \citep{Fontaine2005carbon,Brassard2007origin,koester2014frequency}. WD\,1622+587 has an effective temperature of 23\,430\,K and when compared to models from \citet{Fontaine2005carbon} and \citet{Brassard2007origin}, it cannot be ruled out that a significant fraction of the carbon abundance in WD\,1622+587 may be due to carbon winds rather than the external accretion of carbon-rich planetary material, and so the carbon abundance is unreliable.

The Bayesian framework is unable to model bodies that have such a low oxygen abundance, and therefore does not provide a good fit to the data.

\subsection{Gaia\,J2100+2122} \label{GaiaJ2100}

% Andy's model: Primitive, some heating required to explain abundance patterns with oxygen upper limit and silicon slight depletion as silicon sinks slower than Mg, Ca slight enhancement as Ca sinks faster than Mg. Fe slightly enhanced. 

\begin{figure}
	\includegraphics[width=\columnwidth]{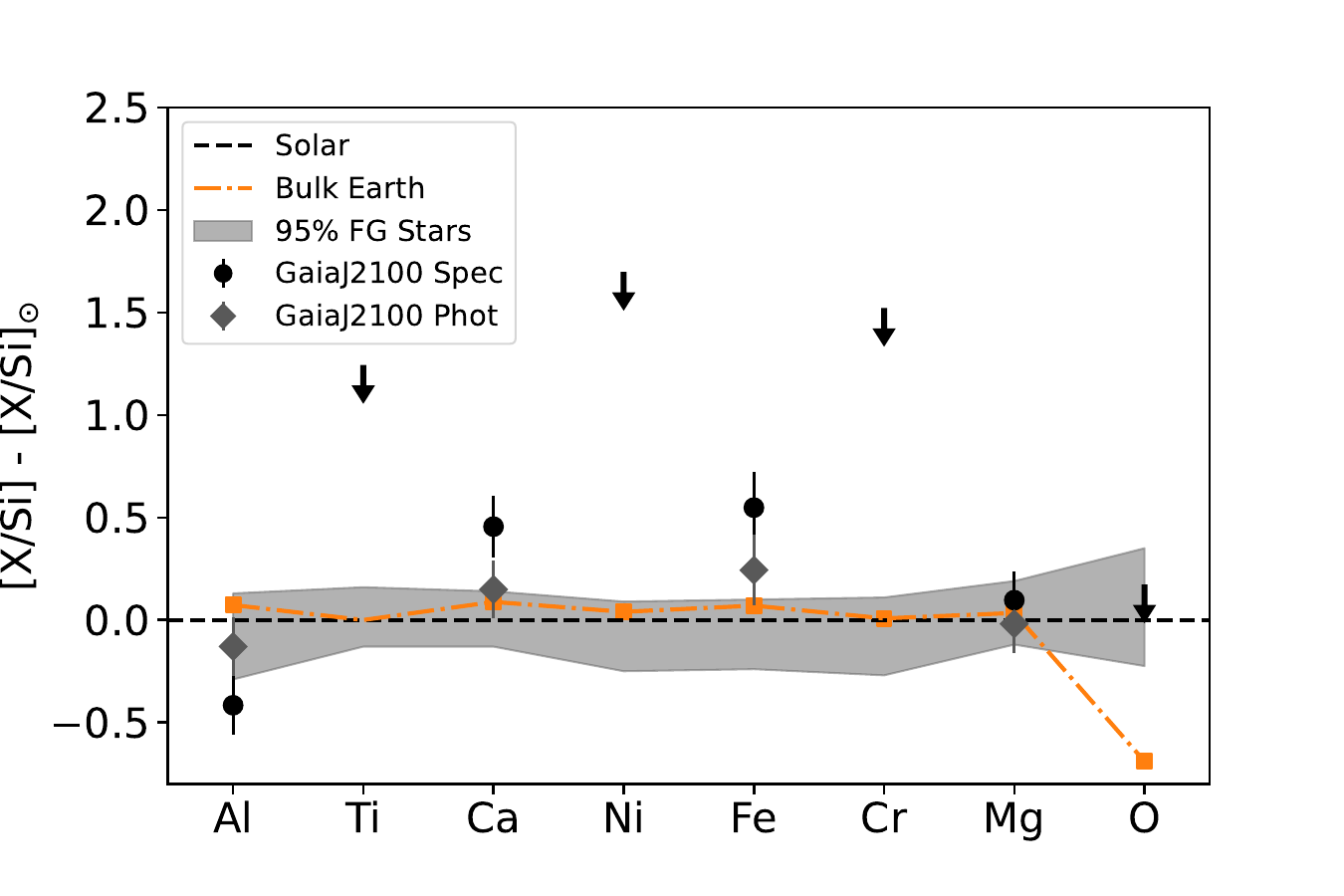}
    \caption{The logarithmic number ratio of elemental abundances in the pollutant material relative to Si, normalised to solar (dashed black line). Observed abundances for Gaia\,J2100+2122 are shown for two different sets of white dwarf parameters, abundances derived using the spectroscopic parameters ($T_\mathrm{eff} = 25570$\,K and $\log(g) =8.10$) shown as black circles, and abundances derived using the photometric white dwarf parameters ($T_\mathrm{eff} = 22000$\,K and $\log(g) =7.92$) are shown as grey diamonds. The grey shaded region shows the 95 percent range of the abundance ratios of FG stars from the Hypatia catalog \citep{hinkel2014stellar}. Upper limits are plotted as arrows and are measured from the base of the arrow.}
    \label{fig:GaiaJ2100-Ab}
\end{figure}

Paper I derives the abundances of the planetary material assuming two different sets of white dwarfs parameters, the spectroscopic white dwarf parameters ($T_\mathrm{eff} = 25570$\,K and $\log(g) =8.10$) and the photometric parameters ($T_\mathrm{eff} = 22000$\,K and $\log(g) =7.92$). Gaia\,J2100+2122 has the largest discrepancies between the two effective temperatures, and so both sets of derived abundances are plotted in Fig.\,\ref{fig:GaiaJ2100-Ab}. For both sets of abundances, the Mg/Si abundance in the exoplanetary body that polluted Gaia\,J2100+2122 reflects bulk Earth, or compositions seen in nearby FG stars. For the abundances derived using the spectroscopic white dwarf parameters, there is an enhancement of Ca and Fe in comparison to solar. The Fe/Si lies 3.2\,$\sigma$ above that of a solar value, when assuming steady state accretion (equation \ref{eq:2}), which may hint at the accretion of a core-rich fragment. However, when considering the abundances derived using the photometric white dwarf parameters, the Fe/Si lies just 1.4\,$\sigma$ above that of a solar value, and the abundance pattern traces that of the compositions seen in nearby FG stars. Although, as reported in Paper I, the abundance ratios are much less sensitive to temperature than the absolute abundances, it is still important to investigate the effect of the white dwarf parameters. In this temperature range (20\,000 -- 30\,000\,K), the more refractory elements (Fe, Ca, Al) transition between dominant ionisation states making the discrepancy between the two sets of abundances more prominent for these elements. 

There is insufficient information from the oxygen upper limit to constrain the volatile content for the body that has been accreted by Gaia\,J2100+2122. Ultraviolet spectroscopy of Gaia\,J2100+2122 should enable this to be possible. 

\textsc{PyllutedWD} invokes a primitive model to explain the abundance patterns. Like with Gaia\,J0611$-$6931, the Fe enhancement is not significant enough to reject the primitive model and invoke the complex core-mantle differentiation model in the context of the Bayesian framework.

\section{Comparing pollution composition in white dwarfs with and without detectable circumstellar gas} \label{pop}

\begin{figure*}
%\captionsetup[subfigure]{font=small} if you like to change caption style
     \begin{subfigure}[b]{0.49\textwidth}
         \includegraphics[width=\columnwidth]{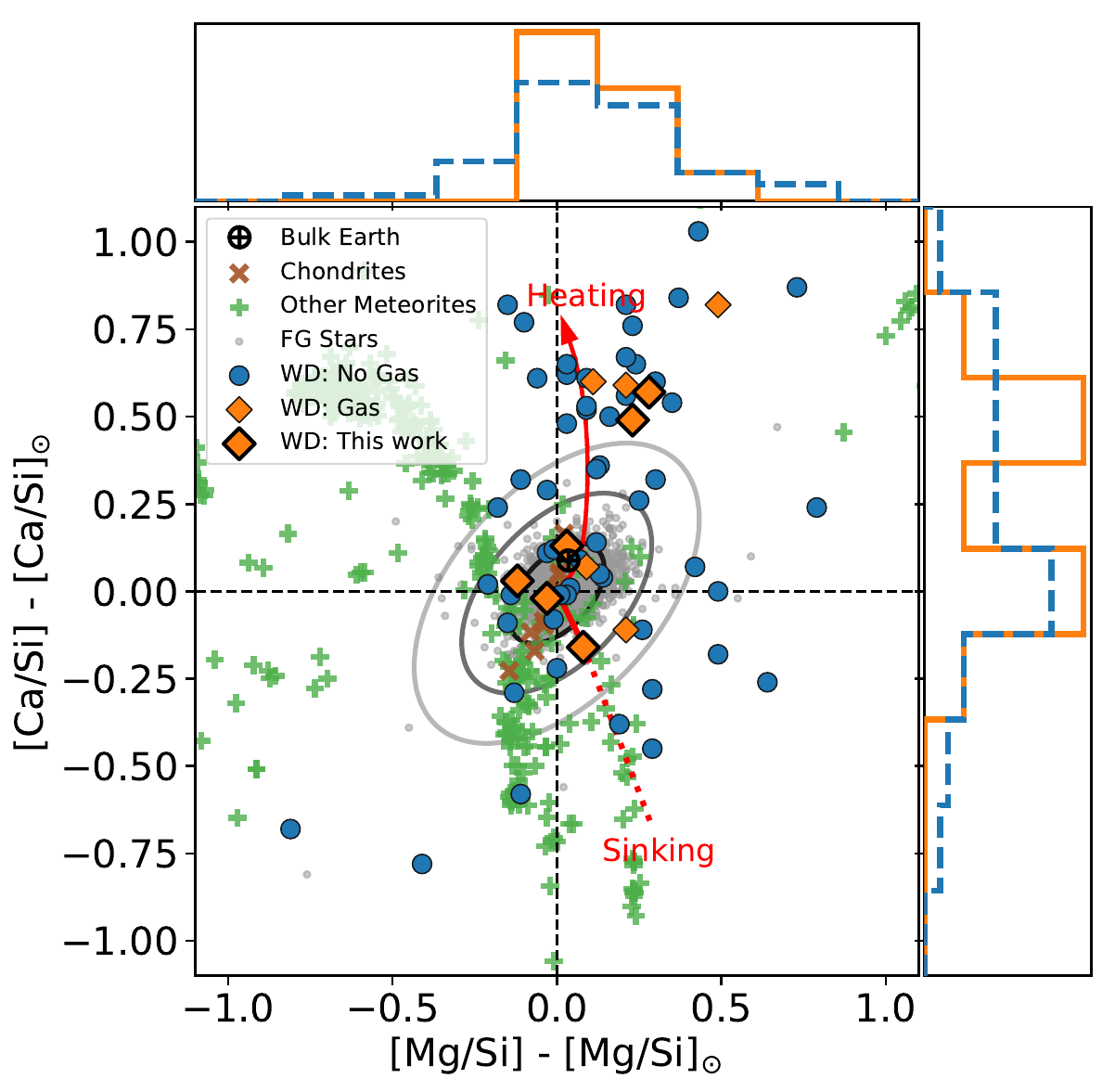}
         \caption[]{} % <---
         \label{fig:ratio-a}
     \end{subfigure}
     \hfill
     \begin{subfigure}[b]{0.49\textwidth}
         \includegraphics[width=\columnwidth]{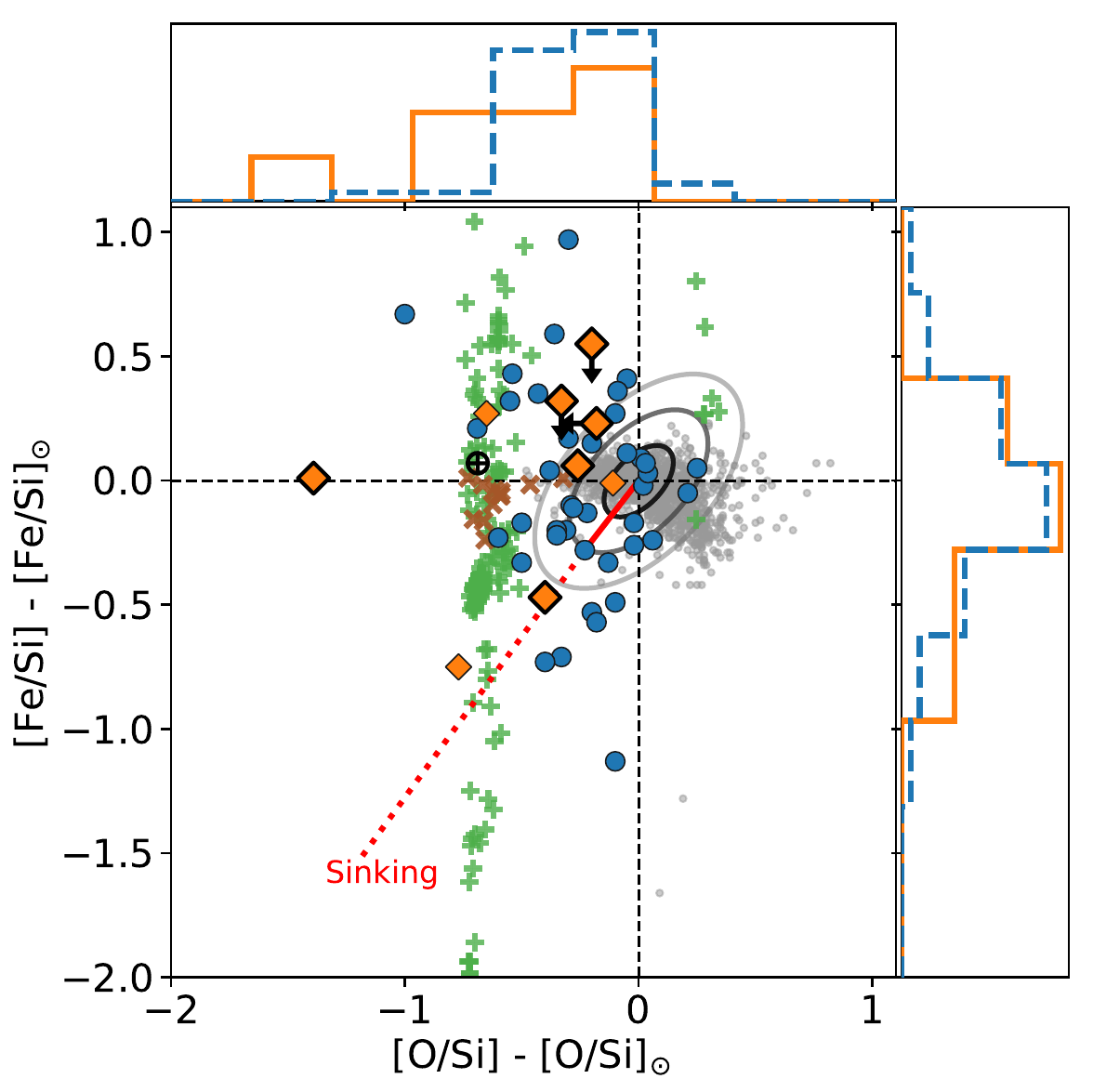}
         \caption[]{} % <---
         \label{fig:ratio-b}
     \end{subfigure}

     \vskip\baselineskip
     \begin{subfigure}[b]{0.49\textwidth}
         \includegraphics[width=\columnwidth]{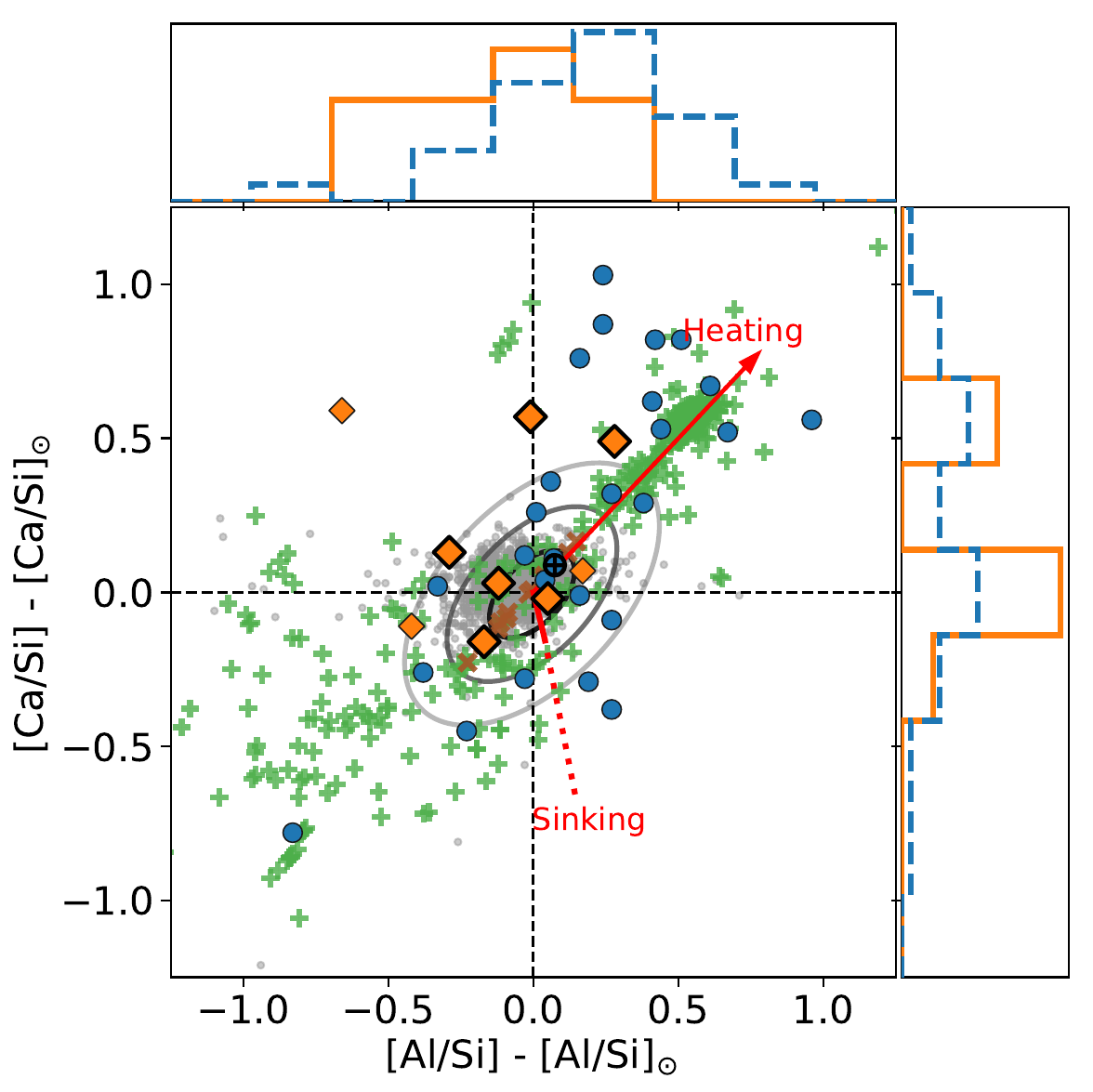}
         \caption[]{} % <---
         \label{fig:ratio-c}
     \end{subfigure}
     \hfill
     \begin{subfigure}[b]{0.49\textwidth}
         \centering
         \includegraphics[width=\columnwidth]{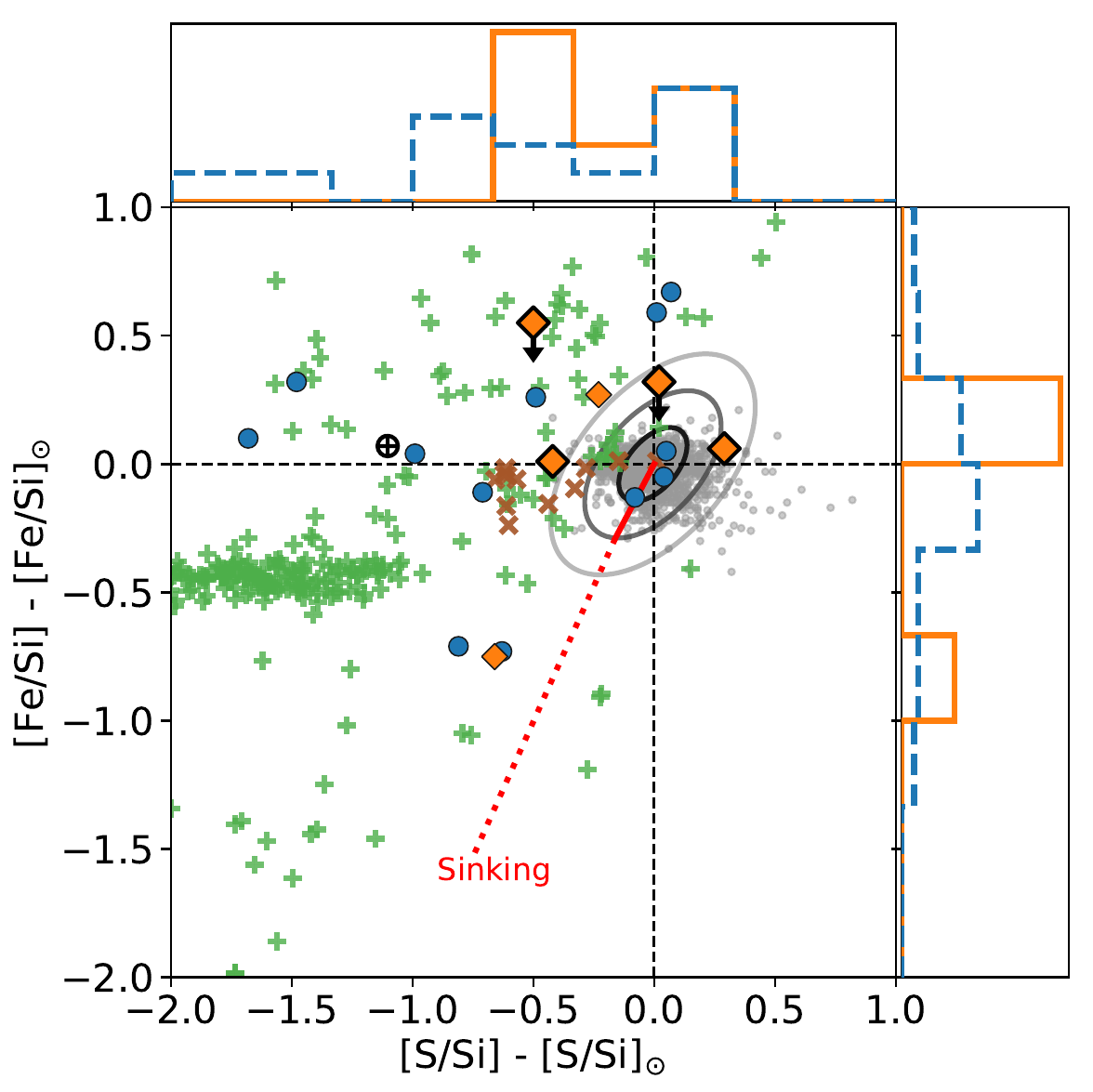}
         \caption[]{} % <---
         \label{fig:ratio-d}
     \end{subfigure}
        \caption{Figures showing the logarithmic number abundance ratio of  polluted white dwarfs comparing those with detectable circumstellar gas in emission (orange) compared to without (blue), normalised to solar (black dashed line) using solar abundances from \citet{Lodders2009solar}. The grey ellipses show a 1\,$\sigma$, 2\,$\sigma$ and 3\,$\sigma$ error ellipse based on solar abundances with individual element errors of 0.1 dex, typical abundance errors for polluted white dwarfs. The arrows show how the abundances are affected by: heating of the pollutant body during formation (see Sections \ref{Bayes} and \ref{pop}) and sinking in the white dwarfs' atmosphere (solid line shows steady state, and dotted line shows 3$\tau _{\mathrm{Si}}$ into the declining phase). The histograms show the 1D distribution for those with detectable circumstellar gas in emission (orange solid line) compared to without (blue dashed line). FG main sequence star data are from \citet{hinkel2014stellar}, chondrite data from \citet{Alexander2019carbonaceous,Alexander2019noncarbonaceous}, and other meteorite (achondrites, stoney-iron meteorites, and iron meteorites) data from \citet{Nittler2004bulk}. Abundances of the polluted material for the white dwarfs with detectable gas discs are from: \citet{dufour2012detailed,melis2012gaseous,wilson2015composition,melis2017differentiated,xu2019compositions, Rogers2023sevenI}, and abundances for white dwarfs without detectable gas discs are from: \citet{zuckerman2007chemical,klein2011rocky,melis2011accretion,zuckerman2011aluminum,gansicke2012chemical,jura2012two,Kawka2012vlt,farihi2013evidence,xu2013two,vennes2013polluted,xu2014elemental,raddi2015likely,farihi2016solar,Kawka2016extreme,gentile2017trace,hollands2017cool,xu2017chemical,blouin2018generation,swan2019interpretation,xu2019compositions,fortin2020modeling,hoskin2020white,Kaiser2021lithium,Gonzalez2021WD1814,klein2021discovery,izquierdo2021gd,Elms2022spectral,Hollands2021alkali,Johnson2022unusual,doyle2023new,izquierdo2023systematic,Swan2023planetesimals,Vennes2024cool}.} % <---
        \label{fig:ratios}
\end{figure*}

The white dwarfs in this sample represent seven of 21 polluted white dwarf systems with detectable circumstellar gas in emission. \citet{xu2019compositions} compared pollution levels for those white dwarfs with and without a detectable circumstellar \textit{dust} disc; no significant difference in abundances of polluting material was found. In this section the study is paralleled, but instead to compare the pollution between white dwarfs with and without detectable circumstellar \textit{gaseous} discs, where without refers to all polluted white dwarfs without any evidence of circumstellar gas in emission.

A Kolmogorov-Smirnov (KS) test was used to assess the null hypothesis that the abundance ratios of material in the photospheres of white dwarfs with detectable gaseous discs in emission and those without originate from the same distribution. The considered abundance ratios assuming the build-up phase included: [Ca/Si], [Mg/Si], [Fe/Si], [O/Si], [Al/Si], [Ni/Si], [S/Si], [P/Si], [Ti/Si], [Cr/Si] where Si was used as a comparison element due to the optical-ultraviolet abundance discrepancy as it is the only element measured in both the optical and ultraviolet. For all abundance ratios, the $p$-values are found to be large, and therefore, the null hypothesis cannot be rejected. Figure \ref{fig:ratios} shows some of these abundance ratios, with normalised histograms that compare those with detectable gaseous discs to those without. It remains plausible that the two samples come from the same distribution, and there is no evidence that the abundance ratios ([Ca/Si], [Mg/Si], [Fe/Si], [O/Si], [Al/Si], [Ni/Si], [S/Si], [P/Si], [Ti/Si], [Cr/Si]) of the material in the photospheres of white dwarfs with gaseous discs are different to those without.

The lithophilic abundances (Si, Ca, Mg, Al) of the planetary material accreted by white dwarfs are compared as shown in Figs.\,\ref{fig:ratio-a} and \ref{fig:ratio-c} with only white dwarfs with detections of Ca, Mg, Si, Al being plotted. The observed abundances in the atmospheres of these white dwarfs as well as white dwarfs from the literature cannot solely be the result of accreting material with a range of compositions similar to the range seen in nearby stars. The range of Ca, Mg, Si, Al abundances seen seems to be influenced by their relative sinking and another process that enhances Ca and Al relative to Si, heating where a portion of the body experienced temperatures higher than 1250\,K. The heating arrow in Figs.\,\ref{fig:ratio-a} and \ref{fig:ratio-c} shows how the composition of a planetesimal can be affected by the ambient temperature during formation in the protoplanetary disc. The arrow shows how as the temperature is increased (or distance from the star is decreased) more volatile elements are unable to condense and so become depleted \citep{Chambers2009analytic}, and a number of white dwarfs appear to be accreting planetary material that follows this trend. Figure \ref{fig:ratio-c} also shows a clear trend in the Solar System meteorites with heating \citep{Nittler2004bulk}, and so it appears that planetary material in exoplanetary systems are going through similar formation conditions and processes as meteorites in the Solar System. There is no significant difference in the abundance ratios of [Ca/Si] or [Al/Si] between those with and without detectable gaseous discs and therefore, there is no correlation with amount of heating and the presence of a gaseous disc. 

\begin{figure*}
\centering
\subfloat[]{
  \includegraphics[width=0.48\textwidth]{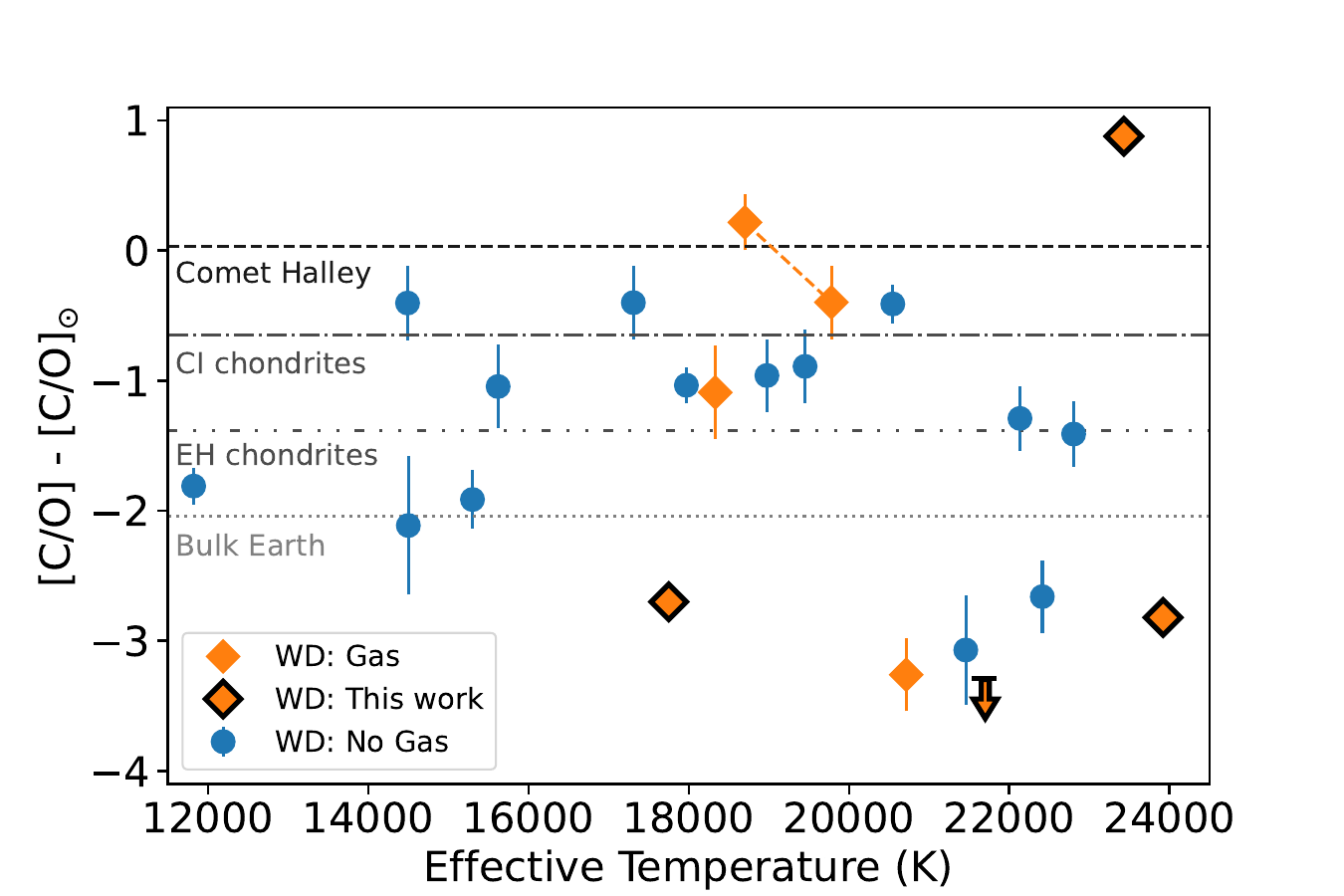}
  \label{fig:CO-plot-a}
}
\hspace{2mm}
\subfloat[]{
  \includegraphics[width=0.48\textwidth]{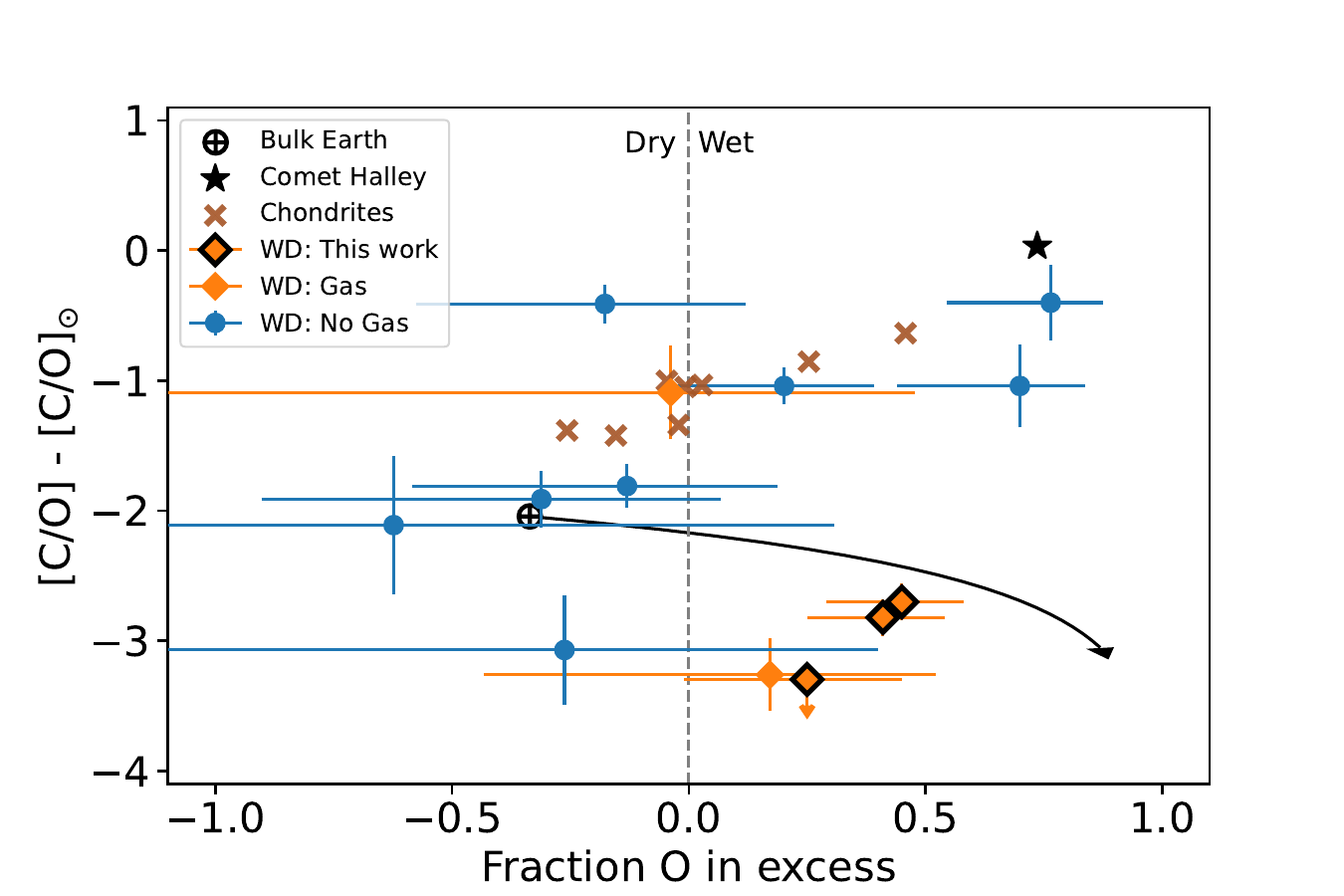}
  \label{fig:CO-plot-b}
}
\caption{(a) [C/O] steady state abundance ratio versus effective temperature of the white dwarf for white dwarfs with (diamonds) versus without (circles) gaseous discs in emission combining data from the literature and this work (Table \ref{tab:WDs-ab-CO}). Both C/O ratios for SDSS\,J0845+2257 are shown connected by a dashed line. (b) [C/O] steady state abundance ratio versus the fraction of oxygen in excess for white dwarfs with both measurements. Any white dwarf pollutants with a fraction of oxygen in excess below -1 are omitted (SDSS\,J0845+2257, WD\,1622+587, and  WD\,0843+516). Bulk Earth, comet Halley, and chondrites are plotted, and the arrow from bulk Earth shows how a planetary body would move if it had the major rock forming abundances of bulk Earth, but with increasing oxygen added.}
\end{figure*}

The [C/O] ratio within a planetary system plays a key role in determining the mineralogy of the resulting planetary material. In an update to the work by \citet{wilson2016carbon}, [C/O] values from the literature along with the four white dwarfs in Paper I with C and O measurements or upper limits, are compiled and presented in Table \ref{tab:WDs-ab-CO} and Figure\,\ref{fig:CO-plot-a}. The [C/O] ratio for polluted white dwarfs with measured C and O abundances plotted against effective temperature is shown in Figure\,\ref{fig:CO-plot-a}, assuming these white dwarfs are in steady state phase. For the warm DAZ white dwarfs, where the sinking timescales are on the order of days to years, the assumption of steady state is reasonable. Similarly, for the DBZs, despite orders of magnitude longer sinking timescales, the presence of circumstellar discs for most of these objects suggests active accretion. These systems are most likely in build-up or steady-state phase and as shown in Table \ref{tab:WDs-ab-CO} the [C/O] ratio changes by $<$\,0.1\,dex so the assumption of steady-state is again reasonable. \citet{wilson2016carbon} finds a bi-modality in the [C/O] ratio with a gap between $-1$ and $-2$. In this work, the observed [C/O] ratios for planetesimals in these exoplanetary systems span from solar to $-3.5$\,dex below solar with no obvious gap, and covers a range in [C/O] comparable to objects in the Solar System. SDSS\,J0845+2257 and WD\,1622+587 exhibit super-solar [C/O] ratios. SDSS\,J0845+2257 has a super-solar [C/O] ratio when analysed with optical data \citep{Jura2015evidence} and sub-solar [C/O] ratio when analysed with ultraviolet data \citep{wilson2015composition}, given this discrepancy it is unknown whether this is truly a carbon rich/oxygen poor planetesimal. As mentioned in Section \ref{WD1622}, the carbon abundance for WD\,1622+587 could potentially contain a contribution from carbon winds, and so it is excluded from the following analysis. Therefore, there is no substantial evidence for any white dwarf accreting carbon rich planetary material. 

Omitting WD\,1622+587, there are five measurements and one upper limit for [C/O] for planetesimals accreted by white dwarfs with circumstellar gaseous discs. The [C/O] ratio of white dwarfs with gaseous discs skews towards lower [C/O] values. As seen in Figs.\,\ref{fig:GaiaJ0006-Ab}, \ref{fig:GaiaJ0510-Ab}, and \ref{fig:GaiaJ0611-Ab}, the three DAZ white dwarfs studied in this work have carbon abundances at or below that of bulk Earth, whereas the oxygen values are close to that of solar, resulting in a low [C/O] ratio. To investigate this further, \ref{fig:CO-plot-b} shows the [C/O] ratio plotted against the fraction of excess oxygen for the white dwarfs reported in Table \ref{tab:WDs-ab-CO} with both measurements. These white dwarfs were selected as they have carbon and oxygen abundances, as well as detections or upper limits for the main rock forming elements hence allowing the fraction of oxygen in excess to be calculated. A positive fraction of excess oxygen indicates leftover oxygen after budgeting, which may have been in the form of water, whilst a negative fraction implies insufficient oxygen, suggesting a more metallic composition is needed to explain the abundance pattern. Comet Halley \citep{lodders1998planetary}, chondrites \citep{Alexander2019carbonaceous,Alexander2019noncarbonaceous}, and bulk Earth \citep{mcdonough2003compositional} are included in the plot, with their oxygen budgeting calculated following the same methodology applied to the white dwarfs for consistency. A positive correlation between the [C/O] ratio and fraction of excess oxygen is evident in Solar System objects and most white dwarfs. Another distinct group of white dwarfs with [C/O] ratios around $-$3 and positive fractions of excess oxygen suggests these white dwarfs have accreted rocky asteroids with additional water. The line from bulk Earth shows how the [C/O] and oxygen excess values would change if additional oxygen was added to the bulk composition; assuming additional oxygen alone is insufficient to explain this trend. All four of the white dwarfs in this region have gas discs in emission, and so it is enticing to speculate that these may be unique, perhaps in an early accretion stage \citep{Brouwers2023AsynchronousII}. However, two white dwarfs with gas emission discs do not fall into this category with one shown and the other with a fraction of oxygen in excess below $-$1. In order to truly understand the correlation between the [C/O] ratio and water in exoplanetary systems, additional white dwarfs with measured abundances of C, O, and the major rock forming elements are required.

\begin{table*}
	\centering
	\footnotesize
	\caption{[C/O] ratios for DA and DB white dwarfs assuming both build up and steady state phase in order of increasing effective temperature. The solar C/O ratio from \citet{Lodders2009solar} is $-0.34$ for reference.}
	\label{tab:WDs-ab-CO}
	\begin{tabular}{lcccrrrrl} % four columns, alignment for each
		\hline
		WD Name & DA/DB & $T_\mathrm{eff}$ & $\log(g)$ & log($\tau _{\textrm{C}}$) & log($\tau _{\textrm{O}}$) & [C/O]$_{\textrm{BU}}$/[C/O]$_{\textrm{SS}}$ & O$_\mathrm{excess}$ & Ref. \\
		  &  &  (K) &  (cm\,s$^{-2}$) &  (yrs) &  (yrs) & & & \\
  \hline	
    
        \vspace{1mm} WD\,2326+049 & DA & 11820 & 8.40 & $-$0.23 & $-$0.48 & $-1.90$/$-2.15\pm0.17$ & $-0.13\substack{+0.32 \\ -0.45}$ & \citet{xu2014elemental} \\
        \vspace{1mm} WD\,1425+540 & DB & 14490 & 7.95 & 6.54 & 6.46 & $-0.67/-0.74\pm0.29$ & $0.76\substack{+0.11 \\ -0.22}$ & \citet{xu2017chemical} \\
        \vspace{1mm} WD\,1145+017 & DB & 14500 & 8.11 & 6.20 & 6.13 & $-2.38/-2.45\pm0.53$$^\dagger$ & $-0.61\substack{+0.92 \\ -2.16}$ & \citet{fortin2020modeling} \\
        \vspace{1mm} WD\,0300$-$013 & DB & 15300 & 8.00 & 6.29 & 6.22 & $-2.18/-2.25\pm0.22$ & $-0.31\substack{+0.38 \\ -0.59}$ & \citet{jura2012two} \\
        \vspace{1mm} WD\,1822+410 & DB & 15620 & 7.93 & 6.37 & 6.30 & $-1.31/-1.38\pm0.32$ & $0.70\substack{+0.14 \\ -0.26}$ & \citet{klein2021discovery} \\
        \vspace{1mm} WD\,2058+181  & DA &  17308 & 7.92 & $-$2.10 & $-$2.27 & $-0.57/-0.74\pm0.28$ & & \citet{wilson2016carbon} \\
        \vspace{1mm} Gaia\,J0611$-$6931 & DA &  17750 & 8.14 & $-2.46$ & $-2.62$ & $-2.87/-3.04\pm0.14$ & $0.45\substack{+0.13 \\ -0.16}$ & Paper I \\
        \vspace{1mm} Gaia\,J2047$-$1259 & DB & 17970 & 8.04 & 5.78 & 5.71 & $-1.30/-1.38\pm0.14$ & $0.20\substack{+0.19 \\ -0.25}$ & \citet{hoskin2020white} \\
        \vspace{1mm} SDSS\,J1043+0855 & DA &  18330 & 8.05 & $-$2.28 & $-$2.46 & $-1.25/-1.43\pm0.36$ & $-0.03\substack{+0.51 \\ -1.30}$ & \citet{melis2016does} \\
        \vspace{1mm} SDSS\,J0845+2257$^*$ & DB & 18700 & 8.00 & 5.73 & 5.66 & $-0.05/-0.12\pm0.21$ & $-1.83\substack{+0.69 \\ -0.90}$ & \citet{Jura2015evidence} \\
        \vspace{1mm} WD\,1953$-$715  & DA &  18975 & 7.96 & $-$2.10 & $-$2.30 & $-1.10/-1.30\pm0.28$ & & \citet{wilson2016carbon} \\
        \vspace{1mm} WD\,1943+163  & DA &  19451 & 7.90  & $-$1.95 & $-$2.18 & $-1.00/-1.23\pm0.28$ & & \citet{wilson2016carbon} \\
        \vspace{1mm} SDSS\,J0845+2257$^*$ & DB & 19780 & 8.18 & 5.29 & 5.20 & $-0.65/-0.74\pm0.28$ & $-1.71\substack{+1.15 \\ -2.04}$ & \citet{wilson2015composition} \\
        \vspace{1mm} WD\,1337+701  & DA &  20546  & 7.95  & $-$1.81  & $-$2.13 & $-0.43/-0.75\pm0.15$ & $-0.17\substack{+0.30 \\ -0.40}$ & \citet{Johnson2022unusual} \\
        \vspace{1mm} SDSS\,J1228+1040 & DA &  20713 & 8.15  & $-$2.17 & $-$2.48 & $-3.30/-3.60\pm0.28$ & $0.18\substack{+0.35 \\ -0.60}$ & \citet{gansicke2012chemical}$^\ddagger$ \\
        %\vspace{1mm} WD\,1536+520 & DB & 20800 & 7.96 & 5.44 & 5.36 & $-0.90\pm0.36$ & $-$0.98 & $0.03\substack{+0.40 \\ -0.67}$ & \citet{farihi2016solar} \\
        \vspace{1mm} WD\,1929+012  & DA & 21457 & 7.90  & $-$1.55 & $-$1.96 & $-3.00/-3.41\pm0.42$ & $-0.27\substack{+0.66 \\ -1.41}$ & \citet{gansicke2012chemical}$^\ddagger$ \\
        \vspace{1mm} WD\,1013+256  & DA &  22133 & 8.02  & $-$1.42 & $-$1.95 & $-1.10/-1.63\pm0.25$ & & \citet{wilson2016carbon} \\
        \vspace{1mm} WD\,0843+516  & DA &  22412 & 7.90 & $-$1.33 & $-$1.84 & $-2.50/-3.00\pm0.28$ & $-3.16\substack{+2.2 \\ -4.7}$ & \citet{gansicke2012chemical}$^\ddagger$\\
        \vspace{1mm} WD\,1647+375  & DA &  22803 & 7.90  & $-$1.24 & $-$1.79 & $-1.20/-1.75\pm0.25$ & & \citet{wilson2016carbon} \\
        \vspace{1mm} WD\,1622+587 & DB & 23430 & 7.90 & 5.07 & 4.97 & $0.64/0.54\pm0.15$ & $-13.8\substack{+3.4 \\ -4.4}$ & Paper I \\
        \vspace{1mm} Gaia\,J0006+2858 & DA & 23920 & 8.04 & $-0.98$ & $-1.72$ & $-2.42/-3.16\pm0.14$ & $0.41\substack{+0.13 \\ -0.16}$ & Paper I \\
		\hline
	\end{tabular}
 \vspace{-3mm}
	\small \begin{flushleft}
    \item \textbf{Notes:} 
    \item $^*$Depending on whether SDSS\,J0845+2257 was analysed using optical or ultraviolet data, the C/O ratio is either super-solar or sub-solar, therefore, given these differences both have been reported. 
    \item $^\dagger$C detection reported as tentative and so error of 0.4\,dex is assumed. 
    \item $^\ddagger$ Updated [C/O] values from \citet{wilson2016carbon} also used.
    
    \end{flushleft}
 
\end{table*}

\section{Discussion} \label{disc}

This paper highlights that white dwarfs with detectable dust and gas discs, including the seven analysed here, are excellent laboratories for studying planetary material as the refractory abundances reflect the broader population of polluted white dwarfs. The analysis presented highlights the range of volatile content in the material accreted by white dwarfs, including three likely accreting material rich in water-ice, one accreting a dry rock, and one accreting an exceptionally dry and oxygen poor body. Additionally, the range of iron content accreted by the white dwarfs studied provides support for the accretion of one mantle-rich and two core-rich fragments of larger differentiated bodies. This section discusses the robustness and implications of these results. Whilst it is possible for observable metal abundances to be seen in white dwarfs in the temperature range studied here (16,000--26,000\,K) due to radiative levitation, the abundances reported are all at least an order of magnitude higher than those predicted by \citet{chayer1995radiative,chayer1995improved}.

\subsection{Consistent refractory planetary abundances in white dwarfs with and without detectable gaseous emission}

There is no difference in the refractory composition of the planetary bodies for those white dwarfs with circumstellar gas in emission compared to those without. However, the analysis is limited by small number statistics as there are small numbers of white dwarfs with gaseous discs in emission and smaller numbers of those with photospheric abundance measurements. Therefore, when comparing abundance ratios between the two samples, for [C/O], [P/Si], [Ti/Si], [Cr/Si] there were five or fewer white dwarfs with gaseous discs with these measurements. Over the coming years surveys such as DESI \citep{DESI2016DESI}, SDSS-V \citep{Kollmeier2017SDSSV}, 4MOST \citep{deJong20194MOST}, and WEAVE \citep{Dalton2012WEAVE} will discover numerous new polluted white dwarfs with circumstellar gaseous discs, allowing these conclusions to be revisited in future studies.

There is tentative evidence that the [C/O] ratio and water content of white dwarfs with gaseous discs are statistically distinct from those without gaseous discs. This may hint that the gaseous disc systems represent the earliest stage in the disruption and accretion of planetary bodies around white dwarfs, with the volatiles accreting early in a gas-rich phase \citep{xu2019compositions} due to the differing sublimation radii for volatile versus refractory elements. If confirmed, this would give new observational insight into how the material tidally disrupts, sublimates, and ultimately accretes on to the white dwarf \citep{Brouwers2023AsynchronousII}, and the methods in which oxygen abundance excesses are interpreted to be the result of water rich bodies should be revisited.

While it is generally expected that all polluted white dwarfs accrete gaseous material due to the extreme temperatures close to the white dwarf, these 21 systems with gas discs in emission have gas that extends outside of the sublimation radius, overlapping with the region where the dust disc resides. However, some white dwarfs may be misclassified as white dwarfs without a gaseous disc. Gas can be hidden below detection thresholds; it is plausible that all systems have gaseous discs but only the top end of the distribution are detectable. Additionally, given the known gas variability \citep[e.g.][]{wilson2014variable}, white dwarfs may have been observed during periods when the emission is weakest or non-existent resulting in a non-detection. Also, a handful of white dwarfs have circumstellar gas in absorption \citep[e.g.][]{fortin2020modeling}. These misclassifications could change the distinction between the sample of white dwarfs with gaseous discs versus those without.

\subsection{The range of volatile contents accreted} \label{Volatiles}

\subsubsection{Water content}

This work shows that there is a wide range in volatile content of the planetary material accreted by the white dwarfs in the sample. Gaia\,J0006+2858, Gaia\,J0510+2315, and Gaia\,J0611$-$6931 have oxygen excesses assuming that the iron is in FeO form to 2.6, 1.0, and 2.8\,$\sigma$ respectively using Monte-Carlo error sampling, and to 2.5, 2.6, and 2.1\,$\sigma$ respectively using the \textsc{PyllutedWD} Bayesian framework. The main difference between the significance of the excess for these two methods is because \textsc{PyllutedWD} samples from the posterior distribution of the elemental abundances for the best fitting model for missing elements, whereas the Monte Carlo method uses the upper limit for the missing elements. None of these systems have a 3\,$\sigma$ detection of an oxygen excess, however, the results do hint towards the accreted bodies being best explained by a combination of rocks and water-ice, adding to the handful of polluted white dwarf systems discovered that have likely accreted water rich bodies \citep{farihi2011possible,farihi2013evidence, raddi2015likely, xu2017chemical, hoskin2020white,klein2021discovery}. The determination of the oxygen excess relies on the assignment of O to Fe, Ca, Al, Si, and Mg, and a number of sources of uncertainty affect this calculation. Whilst for Gaia\,J0006+2858 and Gaia\,J0510+2315 no iron is detected, the upper limit on the potential iron that escaped detection provides a clear minimum to the oxygen excess. The oxygen excesses are significant to at least 2--3\,$\sigma$ considering a highly oxidised form of iron (FeO). If the iron were instead solely in Fe$_2$O$_3$, the oxygen excesses would no longer be as significant, but in reality iron would likely take a mixture of forms, including FeO, Fe$_2$O$_3$, metallic Fe, and FeS. The super-solar abundance of S for Gaia\,J0611$-$6931 points towards a likely contribution of FeS to the composition suggesting that the accreted body is even more water-rich than suggested by the calculation presented in Fig.\,\ref{fig:O-ex}. 

Gaia\,J0644$-$0352 and WD\,1622+587 are accreting dry, rocky bodies, resembling asteroids or terrestrial planets in the Solar System and have no oxygen excess. For these two DB white dwarfs, it is not clear whether the white dwarfs are accreting in build-up or steady state. Regardless of the phase of accretion or oxidation state of the iron, there is no evidence for any excess water in either system. Additionally, Gaia\,J0644$-$0352 and WD\,1622+587 show enhanced levels of [Ca/Si], and Gaia\,J0644$-$0352 shows enhanced levels of [Al/Si]. The enhanced Ca and Al relative to Si might be due to the depletion of Si and Mg-rich minerals in favour of highly refractory-rich minerals, containing Ca, Al, Ti, that would occur if a portion of the body experienced temperatures higher than 1250\,K during formation, seen only in the Solar System as calcium-aluminium-rich inclusions within meteorites. Therefore, it is highly likely that the bodies being accreted by Gaia\,J0644$-$0352 and WD\,1622+587 are truly oxygen-poor and dry. The hotter conditions found in the inner region of the planet-forming discs around the more massive stars, that tend to be the progenitors to white dwarfs, may be better suited to condensing grains with such compositions than the Sun.

Gaia\,J0644$-$0352 and WD\,1622+587 are accreting dry, volatile depleted material, however, trace hydrogen adds mystery to the story. One origin of trace hydrogen is the accretion of water-rich planetary material which could have occurred at any point in its history as the hydrogen would remain on the surface of the white dwarf. Alternatively, hydrogen can be accreted from the interstellar medium (ISM) or be primordial and remain after stellar evolution \citep{gentile2017trace}. If this hydrogen is truly from the accretion of a volatile-rich body in its past then the current abundance is conflicting. It is highly unlikely that the current pollutant material for Gaia\,J0644$-$0352 and WD\,1622+587 contains significant water as oxygen budgeting reveals that all oxygen can be in the form of metal oxides (see Fig.\,\ref{fig:O-ex}). This is a robust conclusion as all major rock-forming elements are detected for both white dwarfs, and the phase of accretion is found to not affect the oxygen excess. If these white dwarfs did accrete a water-rich planetesimal, it did so when it was much younger, with the hydrogen remaining on the surface of the white dwarf. As Gaia\,J0644$-$0352 and WD\,1622+587 are currently accreting volatile depleted material inferred from the heavy elements, but in the past \textit{may} have accreted volatile rich material, this implies that it may be accreting material from a wide range of radial locations, or there is sufficient mixing that in a particular reservoir there are both volatile rich and poor planetesimals. This has implications for dynamical models used to explain how planetesimals form, evolve, and ultimately end up accreted by the white dwarf.

As evidenced in previous work, and reinforced here, volatiles can survive stellar evolution to the white dwarf phase and ultimately accrete onto the white dwarf \citep{jura2010survival,Malamud2016Post}. The range of volatile content accreted by this sample alone shows that white dwarfs accrete material from a wide range of radial locations; dry bodies from the inner regions of planetary systems and wet bodies from the outer regions. White dwarfs likely accreted leftover planetary building blocks; those bodies thought to be key in determining the volatile inventory of planets. For Earth, whilst the debate continues as to exactly when Earth's volatiles arrived, the accretion of material from volatile-rich reservoirs further out in the Solar System played a key role in our planet's habitability \citep{wood2008core,rubie2011heterogeneous,morbidelli2000source, morbidelli2012building,raymond2017origin}. Discovery of water-rich planetesimals in exo-planetary systems is crucial as it informs us that volatile rich bodies can survive formation, and subsequent evolution, and can deliver volatiles to planetary systems as late as in the white dwarf phase.

\subsubsection{Carbon} \label{discussion:carbon}

In this study, no polluted white dwarfs have been found to definitively be accreting material with [C/O] greater than zero, agreeing with the conclusions made by \citet{wilson2016carbon} that there are no carbon rich planetesimals. The carbon to oxygen ratio is crucial for planet formation; if the [C/O] is less than zero, it is expected that oxides and silicates will form and dominate the protoplanetary disc mineralogy, however, if [C/O] is greater than zero then the mineralogy is expected to be dominated by carbides. [C/O] ratios of planetary material greater than zero could be due to inheritance from a carbon rich star, the formation location in the protoplanetary disc and transport mechanisms of the volatiles \citep{Bond2010compositional,Madhu2011carbon,Mordasini2016imprint}. Given the results here, planet formation models focusing on super-solar C/O ratios are unlikely to be required.

\subsubsection{Sulfur}

Sulfur is an abundant element in the universe and is chemically versatile; under solar nebula conditions sulfur tends to be found in H$_2$S in the gaseous phase, and post-condensation in FeS. \citep{pasek2005sulfur}. This work highlights how white dwarfs can provide powerful conclusions regarding the availability of sulfur in planetary bodies. Gaia\,J0611$-$6931 has the highest sulfur abundance observed among polluted white dwarfs as shown in Fig.\,\ref{fig:ratio-d}. There is a correlation between high sulfur and high iron (relative to solar) in the following white dwarfs: Gaia\,J0510+2315, Gaia\,J0611$-$6931, GD378, WD1425+540, PG0843+516, and G238-44. The best explanation for this correlation is that sulfur in the form of iron sulfide (FeS) is prevalent in planetary bodies, as seen in many meteorites and expected for planet-forming discs \citep{kama2019abundant}. For white dwarfs with lots of oxygen, such as Gaia\,J0611$-$6931, it is plausible that their compositions are further dominated by water than predicted from the oxygen excess calculations in Section \ref{Oxygen}, as the Fe will be in the form of FeS rather than FeO. The range of sulfur in both cometary and rocky planetary bodies, as witnessed by these white dwarfs, tells a complex story beyond the scope of this work.

\subsection{Core-mantle differentiation in exoplanetary systems}

The high (low) Fe abundances relative to lithophile species (Ca, Mg, Si) are best explained by the accretion of a fragment that is dominated by core (mantle) material, potentially produced in a destructive collision of a larger body. Within the sample there is evidence for iron depletion for Gaia\,J0644$-$0352 which may indicate a mantle rich fragment has been accreted, and there is tentative evidence for enhanced iron abundances in Gaia\,J0611$-$6931 and Gaia\,J2100+2122 which may be associated with the accretion of core rich fragments. This provides support that core-mantle differentiation in exoplanetary systems occurs and is important, as are the large destructive collisions required in these planetary systems to produce the observed fragments. Planetesimal belts are collisional systems, as witnessed by debris disc observations \citep[e.g.][]{wyatt2021debris}. This implies that during the post main sequence phase either massive planetesimal belts must be present, or there are high levels of collisional excitation, or both.

For Gaia\,J2100+2122 and Gaia\,J0611$-$6931, these conclusions are based on the enhanced iron in comparison to solar abundance with abundances being inconsistent with solar at 3.2 and 2.3\,$\sigma$ respectively. Given this, it is not possible to conclude that Gaia\,J0611$-$6931 is truly accreting iron rich material. For Gaia\,J2100+2122, it is important to note that only one iron line was detected in the optical data. Additionally, throughout this study, the abundances derived using spectroscopic white dwarf parameters have been used to make inferences about the composition of the parent bodies that were accreted onto each white dwarf. Cross-verification using photometric white dwarf parameters found no substantial differences in interpretation for the white dwarfs, except for Gaia\,J2100+2122. In this case, the significance of the iron excess above solar reduces to 1.4\,$\sigma$ and it is not possible to conclude that Gaia\,J2100+2122 is truly accreting iron rich material. Ultraviolet spectroscopy of Gaia\,J2100+2122 will help to break this degeneracy such that more accurate white dwarf parameters and hence planetary abundances can be inferred. 

For Gaia\,J0644$-$0352, the iron abundance is depleted by $>$\,3\,$\sigma$, and the upper limit for Ni is also sub-solar, consistent with the accretion of a fragment of mantle-rich material. Additionally, \textsc{PyllutedWD} finds that the best fit to the data is a model which invokes the accretion of a dry, mantle rich fragment. This work does not consider the possibility that the material is accreting in declining phase (equation \ref{eq:3}), as the fact that there is circumstellar dust and gas is assumed to be evidence of ongoing accretion. However, it should be noted that the Fe depletion in Gaia\,J0644$-$0352 could be explained by accretion in the declining phase rather than mantle-rich material. \textsc{PyllutedWD} assumes no prior information on the accretion phase and does find that declining phase may explain the abundances, however, given that it has a circumstellar disc this is assumed unlikely. 

In the Solar System, large-scale melting in asteroids is fueled by the decay of short-lived radioactive nuclides including $^{26}$Al and $^{60}$Fe, as they are not massive enough to generate heat via gravitational potential energy \citep{urey1955cosmic,elkins2011chondrites,Eatson2024Devolatilization}. The budget of short-lived radioactive nuclides across exoplanetary systems is unclear, with estimates ranging from Solar System levels of $^{26}$Al being very rare, accounting for approximately 1 percent of systems \citep{gounelle2015abundance}, to Solar System levels of $^{26}$Al being common \citep{young2016bayes}. If the bodies accreted by white dwarfs are truly asteroid size, this would provide evidence that planetary systems enriched in short-lived radioactive nuclides are common. Conflictingly, Gaia\,J0644$-$0352 may be accreting a fragment of a parent body that is planet sized and therefore, no short-lived radioactive nuclides would be required to differentiate this parent body. However, further work is required to reduce the errors on the planetary abundances accreting onto Gaia\,J0644$-$0352 needed to provide tighter contains on the parent body pressure and oxygen fugacity at which it differentiated.

\section{Conclusions} \label{Conclusions}

This paper presents an analysis of the abundances of seven polluted white dwarfs reported in Paper I that host both detectable circumstellar gas and dust discs. The observations inform processes that determine the composition of exoplanetary material, including volatile loss, heating, and core formation. The key findings are as follows:

\begin{itemize}    

    \item Diversity in the volatile content of the accreted exoplanetary material is observed, with some white dwarfs studied likely accreting icy, water-rich bodies (Gaia\,J0006+2858, Gaia\,J0510+2315, and Gaia\,J0611$-$6931), some accreting dry, rocky bodies (Gaia\,J0644$-$0352) and one white dwarf (WD\,1622+587) accreting planetary material that is very dry and oxygen poor, opening questions as to the type of chemistry that might exist in exoplanetary systems. Planetary bodies can be perturbed on to white dwarfs from a wide range of locations across exoplanetary systems, with no particular preference for close-in orbits, and in the same system white dwarfs could accrete both dry and wet material. Measuring the volatile budget, most notably water, of exoplanetary material is key to understanding the physical conditions within and on the surface of planets and crucially, their potential habitability.

    \item Gaia\,J0611$-$6931 has a high oxygen abundance in comparison to bulk Earth. Oxygen budgeting revealed it has likely accreted a water-rich body, potentially originating from the outer regions of the surviving planetary system. It also has the highest sulfur abundance detected for a polluted white dwarf, and likely accreted a body enriched with FeS and water.

    \item High Al/Si and/or Ca/Si abundances in two white dwarfs (Gaia\,J0644$-$0352 and WD\,1622+587) suggest they are dominated by highly refractory material, processed at high temperatures. The accretion of highly refractory (Ca, Al-rich) material provides a plausible explanation to explain the high Ca/Si and Al/Si of some white dwarfs.

    \item The depleted iron abundance in comparison to elements such as Si, Mg and Ca for Gaia\,J0644$-$0352 suggests that it accreted a fragment of a mantle rich planetesimal. The pre-fragmentation parent body may have been similar in size to a planet, implying that the bodies polluting white dwarfs, often assumed to be small asteroid sized bodies, did not form small but rather could have been fragments of a larger body. There is tentative evidence for two white dwarfs accreting core-rich fragments due to the enhancement of Fe compared to elements such as Si, and Mg, however, further observations would be needed to confirm this. 

    \item White dwarfs with circumstellar gas are great targets for studying abundances of exoplanetary material. They accrete at high levels so have many elements detected in their spectra and their refractory abundances reflect the general population of polluted white dwarfs. There is tentative evidence that white dwarfs with circumstellar gas in emission are accreting more water-rich material implying that volatiles may accrete earlier in a gas-rich phase. With plenty of data coming out of large optical ground-based surveys over the coming years, effort should be made to prioritise the analysis of new white dwarfs discovered with circumstellar gaseous discs. 
   
\end{itemize}

\section*{Acknowledgements}

We thank the referee for their helpful comments and suggestions which improved the manuscript. AB and LKR acknowledge support of a Royal Society University Research Fellowship, URF\textbackslash R1\textbackslash 211421. LKR acknowledges support of an ESA Co-Sponsored Research Agreement No. 4000138341/22/NL/GLC/my = Tracing the Geology of Exoplanets. SX is supported by the international Gemini Observatory, a program of NSF's NOIRLab, which is managed by the Association of Universities for Research in Astronomy (AURA) under a cooperative agreement with the National Science Foundation, on behalf of the Gemini partnership of Argentina, Brazil, Canada, Chile, the Republic of Korea, and the United States of America. BK acknowledges support from NASA/Keck research contracts 1654589, 1659075, and 1665572. AMB is grateful for the support of a PhD studentship funded by a Royal Society Enhancement Award, RGF\textbackslash EA\textbackslash 180174. STH is funded by the Science and Technology Facilities Council grant ST/S000623/1. C.M.\ acknowledge support from US National Science Foundation grants SPG-1826583 and SPG-1826550. The authors wish to acknowledge Marc Brouwers and Andrew Swan for useful discussions which helped shape the paper. This research is based on observations made with the NASA/ESA Hubble Space Telescope obtained from the Space Telescope Science Institute, which is operated by the Association of Universities for Research in Astronomy, Inc., under NASA contract NAS 5-26555. These observations are associated with programmes 16204 and 16752.

%%%%%%%%%%%%%%%%%%%%%%%%%%%%%%%%%%%%%%%%%%%%%%%%%%
\section*{Data Availability}
This work made use of the code: \textsc{PyllutedWD} which is available on Github at: \url{https://github.com/andrewmbuchan4/PyllutedWD_Public}.

%The inclusion of a Data Availability Statement is a requirement for articles published in MNRAS. Data Availability Statements provide a standardised format for readers to understand the availability of data underlying the research results described in the article. The statement may refer to original data generated in the course of the study or to third-party data analysed in the article. The statement should describe and provide means of access, where possible, by linking to the data or providing the required accession numbers for the relevant databases or DOIs.



\bibliographystyle{mnras}
\bibliography{Master-Bib} % if your bibtex file is called example.bib

%\begin{thebibliography}{99}
%\end{thebibliography}

%%%%%%%%%%%%%%%%%%%%%%%%%%%%%%%%%%%%%%%%%%%%%%%%%%

%%%%%%%%%%%%%%%%% APPENDICES %%%%%%%%%%%%%%%%%%%%%

\appendix

\section{Sinking Timescales} \label{sinking}

Throughout the manuscript, the sinking timescales used are interpolations of grids from \citet{koester2009accretion}, with updated values reported on: \url{http://www1.astrophysik.uni-kiel.de/~koester/astrophysics/astrophysics.html}. The values are reported in Table \ref{tab:sinking-timescales}.

\begin{table*}
	\centering
	\footnotesize
	\caption{Sinking timescales in years used throughout the manuscript assuming the spectroscopic effective temperature and $\log(g)$ from Paper I. 
	% Al, Ti, Ca, Ni, Fe, Cr, Si, Na, O
	}
	\label{tab:sinking-timescales}
	\begin{tabular}{lcccccccccccc} % four columns, alignment for each
		\hline
        WD Name & log($\tau_{\textrm{C}}$) & log($\tau_{\textrm{O}}$) & log($\tau_{\textrm{S}}$) & log($\tau_{\textrm{P}}$) & log($\tau_{\textrm{Mg}}$) & log($\tau_{\textrm{Al}}$) & log($\tau_{\textrm{Si}}$) & log($\tau_{\textrm{Ca}}$) & log($\tau_{\textrm{Ti}}$) & log($\tau_{\textrm{Cr}}$) & log($\tau_{\textrm{Fe}}$) & log($\tau_{\textrm{Ni}}$) \\
		\hline
        Gaia\,J0006 & $-$0.98 & $-$1.72 & $-$1.11 & $-$1.32 & $-$1.43 & $-$1.28 & $-$1.38 & $-$1.69 & $-$1.59 & $-$1.68 & $-$1.70 & $-$1.75 \\ 
        Gaia\,J0347 & $-$1.79 & $-$2.23 & $-$1.96 & $-$1.98 & $-$1.93 & $-$1.89 & $-$1.96 & $-$2.16 & $-$2.16 & $-$2.22 & $-$2.26 & $-$2.30 \\
        Gaia\,J0510 & $-$2.22 & $-$2.57 & $-$2.41 & $-$2.37 & $-$2.26 & $-$2.27 & $-$2.32 & $-$2.49 & $-$2.52 & $-$2.58 & $-$2.61 & $-$2.65 \\
        Gaia\,J0611 & $-$2.46 & $-$2.62 & $-$2.75 & $-$2.59 & $-$2.32 & $-$2.41 & $-$2.45 & $-$2.53 & $-$2.60 & $-$2.66 & $-$2.71 & $-$2.76 \\
        Gaia\,J0644 & 5.49 & 5.41 & 5.22 & 5.21 & 5.29 & 5.25 & 5.26 & 5.16 & 5.07 & 5.05 & 5.02 & 5.01 \\
        WD\,1622 & 5.19 & 5.07 & 4.86 & 4.85 & 4.92 & 4.88 & 4.89 & 4.79 & 4.68 & 4.68 & 4.64 & 4.66 \\ 
        Gaia\,J2100 & $-$0.93 & $-$1.75 & $-$1.05 & $-$1.31 & $-$1.47 & $-$1.28 & $-$1.40 & $-$1.73 & $-$1.61 & $-$1.70 & $-$1.72 & $-$1.78 \\
        \hline
	\end{tabular}
\end{table*}

%%%%%%%%%%%%%%%%%%%%%%%%%%%%%%%%%%%%%%%%%%%%%%%%%%

% Don't change these lines
\bsp	% typesetting comment
\label{lastpage}
\end{document}